\documentclass[fleqn,usenatbib]{mnras}
\usepackage[T1]{fontenc}
\usepackage{ae,aecompl}
\usepackage{graphicx}	
\usepackage{amsmath}	
\usepackage{amssymb}	
\usepackage{tikz}
\usepackage{times}
\usepackage[normalem]{ulem}
\usepackage{multirow}
\usepackage{tabularx}
\usepackage{bm}            
\usepackage{url}
\usepackage{color,colortbl}
\usepackage[dvipsnames]{xcolor}
\usepackage{placeins}
\usepackage{xspace}
\usepackage{rotating}
\usepackage{float}
\usepackage{caption}

\usepackage{hyperref}

\pdfoutput=1

\newcommand{\el}[2]{$^{#1}$#2}
\newcommand{\Ni}{\el{56}{Ni}}
\newcommand{\Msun}{$M_{\odot}$}
\newcommand{\kms}{\ensuremath{~\mathrm{km~s^{-1}}}}

\newcommand{\CIsinglet}{[\ion{C}{I}] $\lambda$8727}
\newcommand{\CIdoublet}{[\ion{C}{I}] $\lambda\lambda$9824, 9850}

\newcommand{\LCIsinglet}{$L_{\text{[C I] $\lambda$8727}}$}
\newcommand{\LCIdoublet}{$L_{\text{[C I] $\lambda\lambda$9824, 9850}}$}

\newcommand{\CaNIRtriplet}{Ca II NIR triplet}
\newcommand{\CaIIdoublet}{[\ion{Ca}{II}] $\lambda\lambda$7291, 7323}

\newcommand{\Mcarbon}{$M_{\text{C }}$}
\newcommand{\McarbonI}{$M_{\text{C I}}$}

\newcommand{\NIIdoublet}{[\ion{N}{II}] $\lambda\lambda$6548, 6583}

\newcommand{\OIsinglet}{[\ion{O}{I}] $\lambda$5577}
\newcommand{\OIdoublet}{[\ion{O}{I}] $\lambda\lambda$6300, 6364}

\newcommand{\OIeightfourfoursix}{\ion{O}{I} $\lambda$8446}
\newcommand{\MgIeighteightosix}{\ion{Mg}{I} $\lambda$8806}

\newcommand{\CIdoubletdiag}{$f_{\text{[\ion{C}{I}]}\lambda\lambda9824, 9850}$}
\newcommand{\CIsingletdiag}{$f_{\text{[\ion{C}{I}]}\lambda8727}$}

\newcommand{\twelveCToSixteenO}{$^{12}$C($\alpha,\gamma$)$^{16}$O}

\newcommand{\Mi}{$M_{\text{He,i}}$}

\newcommand{\codename}{CaNARY}

\newcommand{\fluxunit}{\ensuremath{\mathrm{erg\ s^{-1}\ cm^{-2}\ \text{\AA}^{-1}}}}

\newcommand{\MCOcore}{$M_{\text{CO}}$}
\newcommand{\XtwelveC}{$X_{\text{12C}} $}

\newcommand{\Mzams}{$M_{\text{ZAMS}}$}

\newcommand{\MpreSN}{$M_{\text{preSN}}$}

\newcommand{\SUMOcodename}{\texttt{SUMO}}

\newcommand{\updated}[1]{\textcolor{black}{{#1}}}
\newcommand{\ajc}[1]{\textcolor{black}{{#1}}}

\definecolor{Gray}{gray}{0.9}

\usepackage{color}
\definecolor{mycolor}{rgb}{.0,.3,1.}

\begin{document}



\title[Carbon lines in Stripped Envelope SNe]{Formation and Diagnostic Use of Carbon Lines in Stripped-Envelope Supernovae }

\author[Barmentloo \& Jerkstrand]
{\parbox{\textwidth}{
    Stan Barmentloo$^{1}$ \thanks{E-mail: stan.barmentloo@astro.su.se},
    Anders Jerkstrand$^{1}$  \thanks{E-mail: anders.jerkstrand@astro.su.se}
    }
\vspace{.2cm}\\
$^{1}$ The Oskar Klein Centre, Department of Astronomy, Stockholm University, AlbaNova, SE-10691 Stockholm, Sweden
}


\pubyear{2025}

\label{firstpage}
\pagerange{\pageref{firstpage}--\pageref{lastpage}}
\maketitle


\begin{abstract}
{
Carbon is one of the main end products of nucleosynthesis in massive stars. In this work, we study the emission signatures of carbon in spectra of stripped envelope supernovae (SESNe). A grid of model nebular spectra is created using the NLTE radiative transfer code \SUMOcodename{}, with stellar evolution- and explosion models as inputs. In the models, \CIsinglet{} and \CIdoublet{} are identified as the only significant optical carbon lines, with contribution from both the O/C and He/C zones. To obtain estimates of \LCIsinglet{}, \ajc{which is blended with the Ca II triplet}, we introduce and apply the \codename{} code, a publicly available \ajc{Monte Carlo scattering code}. We study carbon lines in a sample of SESNe, and find that luminosities of \CIdoublet{} relative to the optical spectrum increase with time, just as in our model grid. However, the relative luminosities of both \CIdoublet{} and \CIsinglet{} are overproduced in our models. \updated{Multiple explanations for this discrepancy, such as too high carbon abundances in the stellar evolution models and underestimated cooling through molecule formation, are investigated. For those SNe where both lines are clearly observed, we use an analytical formalism to constrain their ejected carbon masses to the range $\sim$ 0.2 -- 2 \Msun{}. However, several SNe yield upper limits of 0.05 \Msun{}.} We also show that \CIdoublet{} is a useful line to diagnose both carbon mass and the extent of the He/C zone. We strongly encourage observers and instrumentalists to target \CIdoublet{} in future SN observing campaigns.
}

\end{abstract}


\section{Introduction}
\label{sec:introduction}

Once they run out of nuclear fuel, massive stars ($M_{\rm {ZAMS}} \gtrsim$ 8 \Msun{}) end their lives as a core-collapse supernova (CCSN) \citep{Ekström_2025_Massivestarreview, Jerkstrand_2025_CCSNereview}. In this explosion, these stars eject large amounts of heavy elements created through nucleosynthesis in their cores. In this way, CCSNe are the main astrophysical source for the majority of elements with $Z \lesssim 36$ \citep{Timmes_1995_Canon, Kobayashi_2020_Canon, Sieverding_2023_3DCCSNeyields}.

One of the elements to which CCSNe contribute significantly is carbon. However, there still remains uncertainty as to the relative importance of massive stars (\ajc{through} CCSNe) on the one hand, and low- and intermediate mass stars (\ajc{through} AGB-stars) on the other \citep{Franchini_2020_CarbonOrigins, Farmer_2021_Carbonintroduction}. The \ajc{relative importance of} these contributions, and thus when and where most carbon is formed, has important consequences for the star formation history and IMF of galaxies \citep{Matteucci_2012_Starformation, Jerabkova_2018_galaxywideIMF, Romano_2020_carboningalaxies}. Obtaining better constraints on carbon yields from CCSNe will improve our understanding in these areas.

Another reason to study carbon yields in massive stars is the \twelveCToSixteenO{} reaction rate. This reaction directly influences the abundance of \el{12}{C} left at the end of He-burning \citep{Laplace_2025_CNeshellmergers}. This abundance in turn sets the compactness of the star \citep{Chieffi_2020_compactness}, which governs its explodability \citep{Sukhbold_2014_Compactness,Xin_2025_explodability12C16O}. Despite its importance, the \twelveCToSixteenO{} reaction rate has historically been, and today still is, relatively uncertain (for a good overview, see section 2.2 in \citet{Pepper_2022_history12Cto16O}). 

Studying SNe provides a valuable avenue for constraining nucleosynthetic yields in massive stars, as their spectra do not only contain signatures from the surface layers, but also, during the so-called \textit{nebular phase} ($\gtrsim$ 100 -- 150 d post explosion), the interior regions \citep{Heger_2003_Canon, Hillier_2012_SNinterioryields, Jerkstrand_2017_Book}. Using models of the elemental compositions of exploded stars (e.g. \citet{Limongi_2018_SNmodels, Ertl_2020_explosions, Ma_2025_CarbonFromBinaryModels}) as inputs, synthetic nebular spectra can be created with \ajc{spectral formation} codes (e.g. \SUMOcodename{} \citep{Jerkstrand_2011_SUMOa, Jerkstrand_2012_SUMOb, Jerkstrand_2014_SUMOc}, \texttt{CMFGEN} \citep{Hillier_1998_CMFGENa, Dessart_2010_CMFGENb, Hillier_2012_CMFGENc} and \texttt{EXTRASS} \citep{Jerkstrand2020,vanBaal_2023_EXTRASS, vanBaal_2024_EXTRASS2}) that solve for the physical conditions (temperature, excitation- and ionisation structure) and radiative transfer in the nebula. Comparing these synthetic spectra to observed ones in turn allows for estimations of the nucleosynthetic yields \ajc{and progenitor mass estimates. The results can increasingly often be compared with analysis of progenitor star detections} in pre-explosion images \citep{Smartt_2015_Progenitors, VanDyk_2017_Progenitorimaging, vanDyk_2023_Progenitors}. 

While carbon emission has been modeled and studied for individual SNe, such as the Type Ib SN 1985F \citep{Fransson_1989_Canon}, the Type Ic SN 2007gr \citep{Mazzali_2010_2007grcarbon} and three Type IIb SNe \citep{Jerkstrand_2015_Canon}, a comprehensive study of carbon in a large sample of SNe is still lacking. In part, this is due to the difficulties associated with \ajc{observing and} measuring the two significant nebular carbon lines, \CIsinglet{} and \CIdoublet{}. \CIsinglet{} is part of the \CaNIRtriplet{}-complex, a region which is known to host (at least) five more non-negligible lines in nebular SESNe, some of which (e.g. the Ca NIR triplet itself) are optically thick and thus further complicate estimates of line luminosities. On the other hand, \CIdoublet{} is an effectively isolated feature, but its wavelength region is often not covered in spectrographs used for obtaining nebular SESN spectra. When the region is covered, it often suffers from poor S/N, as it is close to the edge of the spectrograph. For ground-based observations, the blue half of \CIdoublet{} also suffers non-negligible atmospheric absorption ($\sim$20\%).  

In this work we attempt to face these challenges and study carbon line formation in a sample of CCSNe. To this end, we analyse a variation of the sample of spectra presented in \citet{Barmentloo_2024_Nitrogen}, which contains 123 nebular spectra of 41 SESNe (Stripped Envelope SNe, i.e. SNe of stars that lost their hydrogen envelope, \ajc{and in some cases also the He envelope}). To better understand the formation of carbon lines for such SNe, we compute a grid of NLTE radiative transfer models using \SUMOcodename{}. 

The paper is built up as follows: in Section \ref{sec:theory}, we start with a short overview of the theory on the formation of the two most important carbon lines in SESNe, \CIsinglet{} and \CIdoublet{}. Then, Section \ref{sec:observations} describes the SESN sample that is studied in the remainder of the work. Section \ref{sec:methodology} covers how the synthetic model spectra were simulated, as well as how we estimated \LCIsinglet{} and \LCIdoublet{} for our sample. These luminosity estimates are presented and compared to the \SUMOcodename{} models in Section \ref{sec:results}, and they are then used to estimate \ajc{the carbon mass} for the observed SNe. Section \ref{sec:discussion} discusses potential explanations for the lack in carbon luminosity observed in the section before. Finally, the paper is closed of with a summary and outlook in Section \ref{sec:conclusion}.

\section{Theory on Carbon Line Formation}
\label{sec:theory}

\begin{figure}
    \centering
    \includegraphics[width=0.475\textwidth]{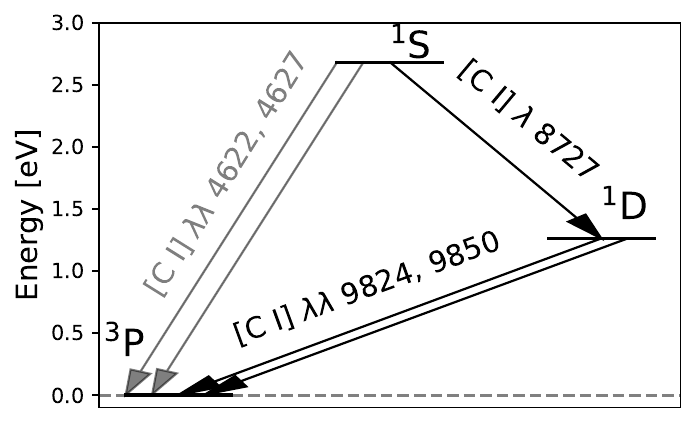}
    \caption{A diagram showcasing the three lowest lying \ajc{terms} in the \ion{C}{I} \ajc{atom}, and optical + NIR transitions connecting them. The [\ion{C}{I}] $\lambda\lambda$4622, 4627 doublet has been shaded in grey, as it has not been identified in SN spectra \ajc{(their A-values are over 100 times smaller than for the 8727 path)}.}
    \label{fig:carbon_level_structure}
\end{figure}

To better understand the formation and use of carbon lines in nebular spectra, it is important to understand under which conditions the different lines form. During the nebular phase, the decreasing densities and temperatures mean that only low lying levels ($\sim$ a few eV) in the atom can be effectively populated and cause strong emission lines. For \ion{C}{I}, these levels are (besides the triplet ground state 2s$^{2}$2p$^{2}$($^3$P) ) the 2s$^{2}$2p$^{2}$($^1$D) level at 1.26 eV and the 2s$^{2}$2p$^{2}$($^1$S) level at 2.68 eV (see Figure \ref{fig:carbon_level_structure}). Of the lines connecting these levels, only \CIsinglet{} and \CIdoublet{} have been robustly observed and modeled (e.g. in \citet{Fransson_1989_Canon, Mazzali_2010_2007grcarbon, Jerkstrand_2015_Canon, Dessart_2023_grid}). The doublet line [\ion{C}{I}] $\lambda \lambda$4622, 4627 has a transition strength $>$ 100 times lower than the competing transition \CIsinglet{}, making this line insignificant. For \ion{C}{II} (which is present in significant quantities throughout the nebular phase) the first excited level resides at 5.33 eV, and no lines from this or any of the higher levels have been observed or found \ajc{to be strong} in spectral models in nebular-phase SESNe. 

Besides their upper levels being effectively populated, \CIsinglet{} and \CIdoublet{} are of particular interest and use for another reason. As these lines are forbidden, their transition rates are low at $A=6 \times 10^{-1}$ s$^{-1}$ and $A=2.9 \times 10^{-4}$ s$^{-1}$, respectively. \ajc{This means that collisional (de-)excitation tends to dominate, bringing about LTE} \updated{for these levels with respect to the ground state}. For such transitions, the luminosity $L$ in the line can be described (in the optically thin case, valid for both our lines; see Table 1 in \citet{Jerkstrand_2017_Book}) using Equation \ref{eq:lte_luminosity} (eq. 42 in \citet{Jerkstrand_2017_Book}):
\begin{equation}
L=M_{ion}\left(\mu m_p\right)^{-1} A h \nu \frac{g_u}{Z(T)} e^{-E_u / k T},
\label{eq:lte_luminosity}
\end{equation}
where $\mu m_{p}$ is the weight of a carbon atom, $E_{u}$ is the energy of the upper level, $g_{u}$ is the statistical weight and $Z(T)$ is the partition function. This equation has only two unknowns ($M_{ion}$ and $T$), meaning that if we can extract \LCIsinglet{} and \LCIdoublet{} from the spectrum, we can isolate $T$ by taking the ratio of these luminosities. In turn, this allows us to calculate $M_{ion}$. We can thus estimate the mass of \ion{C}{I} in the ejecta from a nebular spectrum. As will be discussed in Section \ref{subsec:carbon_mass_estimation}, the ionisation fraction of neutral carbon is not expected to vary much with time or progenitor mass, meaning that an estimate of $M_{\ion{C}{I}}$ can, in turn, be translated to an estimate of \Mcarbon{} without too much \ajc{uncertainty}.

The decrease of temperature and density in the ejecta over time means that eventually, the upper level of \CIsinglet{} is no longer \ajc{dominantly depopulated} by thermal collisions, and the analytical formalism starts to break down. \ajc{This happens for \CIsinglet{} well before for \CIdoublet{} due to its much higher A-value}, \updated{meaning that it has a much higher critical electron density $n^{crit}_{e}$ (2.5 $\times$ 10$^{7}$ and 2.6 $\times$ 10$^{4}$ g cm$^{-3}$ for the $^1$S and $^1$D levels, respectively; see section 3.2 in \citet{Jerkstrand_2017_Book})}. In such an NLTE case, the luminosity in a line is instead given by Equation \ref{eq:nlte_luminosity} (eq. 50 in \citet{Jerkstrand_2017_Book}):
\begin{equation}
L=\left(\mu m_p\right)^{-1} h \nu \times M_f Q_{u f}(T) \frac{g_u}{g_f} e^{-\left(E_u-E_f\right) / k T} n_e,
\label{eq:nlte_luminosity}
\end{equation}
where $Q_{uf}$ the collision rate in cm$^{3}$ s$^{-1}$ between the upper level of the transition $u$ and the feeding level $f$ and $n_{e}$ the electron density. \ajc{For the \ion{C}{I} lines, the main feeding state is always the ground term, which is close to the \ion{C}{I} mass}. In this case there are then three unknowns (\ajc{\McarbonI{},$T$,$n_e$}), so that we can no longer determine \McarbonI{} from just the \ajc{\LCIsinglet{} and \LCIdoublet{}} line ratio. It is thus of the utmost importance to know at what epoch the LTE assumption breaks down. \citet{Jerkstrand_2015_Canon} applied the LTE formalism to the line ratio of \OIsinglet{} / \OIdoublet{}, \ajc{which form analogously to \CIsinglet{} and \CIdoublet{}}, and found that for \OIsinglet{} (which has an upper level of 4.2 eV above the ground state) this occurs around $\sim$ 150d post explosion. If and when this break down occurs for \CIsinglet{} for the models in our grid is investigated in Section \ref{subsec:carbon_mass_estimation}. A further assumption that is made when using Equation \ref{eq:lte_luminosity}, is that \ajc{the carbon emission can be well described by a single zone, single temperature model, an ansatz with unknown accuracy}. 
Again, the impact that this assumption has on our \Mcarbon{} estimates is discussed in Section \ref{subsec:carbon_mass_estimation}.

\section{Observational Data}
\label{sec:observations}

As mentioned in Section \ref{sec:introduction}, the analysis in this work is performed on a variation of the sample of SESNe spectra presented in \citet{Barmentloo_2024_Nitrogen}. The majority of spectra in this sample were obtained through the Weizmann Interactive Supernova Data Repostiory (WISeREP) \citep{Yaron_2012_Wiserep}. A few spectra of the original sample were excluded, as they did not cover the wavelength ranges of \CIsinglet{} and \CIdoublet{}. Some new spectra were added as well, especially to increase the number of spectra covering the \CIdoublet{} region.

For clarity, we have split the sample in this work into two subsamples; one with all spectra considered in this work (called the 'Optical sample') and one with only those spectra that \textit{do} cover the \CIdoublet{} region (called the 'NIR sample'). It is thus only spectra in the NIR sample that can be used to estimate the mass of \ion{C}{I} using the method described in Section \ref{sec:theory}. The Optical and NIR samples are summarised in Tables \ref{tab:optical_sample} and \ref{tab:NIR_sample}, respectively\footnote{All spectra and code used in this work will be made publicly available at \url{https://github.com/StanBarmentloo/carbon_emission_in_SESNe}}. 

To use the spectra in the NIR sample for determining \ion{C}{I} mass, they need to be flux calibrated. As the quality of flux calibrations for spectra on WISeREP was found to vary, we decided to perform our own flux calibrations. Below we describe the additional data obtained to be able to perform these calibrations.

\subsection{Photometry}
\label{subsec:photometry}

The first step in flux calibrating any spectrum is obtaining photometry. Most of the photometry used in this study was obtained through the Open Supernova Catalog (OSC) \footnote{This catalogue can be found at \url{https://github.com/astrocatalogs/supernovae}}. For each SN, the best sampled band was taken and when necessary linearly (in magnitude-space) interpolated to obtain the apparent magnitude at the time of each spectrum. In case no (well sampled) OSN photometry was present, we obtained photometry directly from the literature (sources and filters indicated in fifth and sixth columns of Table \ref{tab:NIR_sample}). For eight SNe (four with NIR sample spectra), no photometry was found in either OSN or the literature, making it impossible to determine \McarbonI{} estimates for those using the LTE formalism from Section \ref{sec:theory} alone (relative luminosity comparisons are still possible, see Sections \ref{sec:results_9850} \& \ref{sec:results_8727}. Finally, the spectra were scaled to the photometry using the python-package \texttt{speclite} \citep{Kirkby_2024_speclite}.

\subsection{Extinction}
\label{subsec:extinction}

None of the photometry above has yet been corrected for extinction. We corrected for extinction by the Milky Way (MW) using the catalogues of \citet{Schlafly_2011_Extinctionmaps}, accessed through the NASA/IPAC Extragalactic Database (NED). Extinction by the host is more difficult to consistently take care of. For many of our objects, authors have used relations between the equivalent width of the \ion{Na}{I} D line and $E(B-V)_{\text{host}}$, and then assumed a reddening law for the host using $R_{V}$ = 3.1. While this is common practice, multiple studies have shown that for the resolutions typical in SN spectra, \ion{Na}{I} D is a poor proxy for extinction \citep{Blondin_2009_NaIDbad, Poznanski_2011_NaIDbad}\footnote{However, for spectra with sufficiently high resolutions, the relations may still hold \citep{Poznianski_2012_highres}}. Furthermore, setting $R_{V}$ to 3.1 assumes the dust in the host galaxy to behave as MW dust, which is not necessarily the case \citep{Stritzinger_2018_hostreddening}. 

A more reliable method for estimating extinctions is using multi-color light curves of minimally reddened SNe 
as templates. \citet{Stritzinger_2018_hostreddening}
were able to obtain estimates for the host extinction $A_{V}$ for a sample of SESNe using this methodology. For their sample, typical $A_{V}$ values ranged between $\sim$ 0.2 -- 1 \ajc{mag}, which would translate to corrections increasing estimated carbon luminosities by 20 -- 150 \%. In case no $A_{V}$ estimate is present for the host, \Mcarbon{} estimates may thus be underestimated with equal percentages. Unfortunately, the sample of \citet{Stritzinger_2018_hostreddening} only obtains extinction values for a single object in our NIR sample, so that for the majority of our sample, low-resolution \ion{Na}{I} D estimates are the only extinction estimates available. 

The total extinctions adopted for our NIR sample are shown in the seventh column of Table \ref{tab:NIR_sample}. We also indicate through which method these extinctions were obtained. It is important to note that only three of these objects have extinctions determined through either high resolution \ion{Na}{I} D measurements or using the method by \citet{Stritzinger_2018_hostreddening}, with some objects having no extinction estimate at all. This will strongly increase the uncertainty on the results for \Mcarbon{} estimates in Section \ref{subsec:carbon_mass_estimation}, but not on the relative luminosities obtained in Sections \ref{sec:results_9850} \& \ref{sec:results_8727}.

\subsection{Distances}
\label{subsec:distances}

Finally, one needs accurate distance estimates for \ajc{absolute luminosity determinations}. As the majority of our sample is not in the Hubble flow, one can not simply take the redshift and adopt a cosmology to calculate these distances. Instead, we obtained most of our distances from the Cosmicflows-3 (CF3) database \citep{Tully_2009_Cosmicflows, Tully_2016_Cosmicflows3, Graziani_2019_Cosmicflows}. In case no CF3-distance was available, we adopted the mean value of the distances available in the NED database. In case no redshift independent distances were available, we obtained the distance through the redshift with a $H_{0} = 67.8$ \kms Mpc$^{-1}$, $\Omega_{m}$ = 0.308, $\Omega_{\Lambda}$ = 0.692 cosmology. The adopted distances and their source are provided in the final two columns in Table \ref{tab:NIR_sample}.

\section{Methodology}
\label{sec:methodology}

\begin{table*}
    \centering
    \begin{tabular}{c|c|c|c|c|c|c|c|c}
        \hline
         model & $M_{\text{ZAMS}}$ & $M_{\text{preSN}}$ & $M_{\text{ej}}$ & $V_{\text{core}}$ & $M_{\text{$^{56}$Ni}}$  & $M_{\text{C, tot}}$
         & $M_{\text{C, O/C}}$ & $M_{\text{C, He/C}}$\\
         \hline\hline
         he3p3 & 16.1 & 2.67 & 1.2 & 4300 & 0.055 & 0.055 (\ajc{4.6\%}) & 0.033 \ajc{(60\%)} & 0.021 \ajc{(38\%)} \\
         he4p0 & 18.1 & 3.15 & 1.62 & 4500 & 0.061 & 0.097 (\ajc{6.0\%}) & 0.060 \ajc{(62\%)} & 0.030 \ajc{(31\%)} \\
         he6p0 & 23.3 & 4.45 & 2.82 & 5700 & 0.084 & 0.25 (\ajc{8.9\%}) & 0.12 \ajc{(48\%)} & 0.12 \ajc{(48\%)}\\
         he8p0 & 27.9 & 5.64 & 3.95 & 4000 & 0.061 & 0.49 (\ajc{12.4\%}) & 0.11 \ajc{(22\%)} & 0.34 \ajc{(69\%)}\\
         \hline
    \end{tabular}
    \caption{A table summarising the main model parameters for the four ejecta models by \citet{Woosley_2019_inputmodels} and \citet{Ertl_2020_explosions} adopted in this work. Masses are in \Msun{}, and $V_{\text{core}}$ is in \kms{}. In the $M_{\text{C, tot}}$ column, percentages refer to the part of $M_{\text{ej}}$ that is in carbon. In the final two columns, percentages refer to the part of $M_{\text{C, tot}}$ that is in this zone. The He/C mass is the combination of the core and envelope parts.}
    \label{tab:main_properties}
\end{table*}

The work necessary to understand carbon line formation in our SESNe sample consists of two main parts. The first part is to create a grid of model spectra to study the physical parameters most relevant for the formation of carbon lines, as well as to compare our \ajc{observational} sample to. This grid was created using \SUMOcodename{} \citep{Jerkstrand_2011_SUMOa, Jerkstrand_2012_SUMOb, Jerkstrand_2014_SUMOc}, which is a Monte Carlo radiative transfer code that solves the NLTE equations for the temperatures and excitation- and ionisation structures within the nebula. The details on how we created this model grid are described in Section \ref{sec:nlte_grid}. 

The second part of our work is to accurately estimate the luminosities of \CIsinglet{} and \CIdoublet{} \ajc{in the data sample}, so that they can be compared to the luminosities in the model grid, as well as to be used to \ajc{directly} estimate the carbon mass in the ejecta \ajc{with the method} described in Section \ref{sec:theory}. The methodology we implement for estimating \LCIdoublet{} is described in Section \ref{subsec:estimating_9850}, and that for \LCIsinglet{} in Section \ref{subsec:estimating_8727}.

\subsection{NLTE Model Grid}
\label{sec:nlte_grid}

In this subsection, we will discuss how the \SUMOcodename{} models in our work were computed, and with which choices of parameters. The synthetic spectra are shown in Figures \ref{fig:spectra_150d} -- \ref{fig:spectra_400d}. The eventual carbon line formation in these models, as well as its sensitivity to these choices, is discussed in Section \ref{sec:results_line_formation}.

\subsubsection{Ejecta models}
\label{subsec:ejecta_models}

The grid of \SUMOcodename{} models created and used in this work should be understood as a variation of the grid used in \citet{Barmentloo_2024_Nitrogen}, and we therefore refer the reader to sections 3.1 and 3.2 in that article for a detailed description of the ejecta models (originating from \citet{Woosley_2019_inputmodels} and \citet{Ertl_2020_explosions}) and hydrodynamical mixing prescription. The he5p0 model that was present in \citet{Barmentloo_2024_Nitrogen} was dropped as its high core- and thus line velocity of 7 000 \kms{} was found to not match any of our observed spectra. A summary of the most important model parameters for our four remaining ejecta models is provided in Table \ref{tab:main_properties}, \updated{whereas detailed compositional tables are presented in Appendix \ref{app:compmodels}}.

\subsubsection{Molecules}
\label{sec:molecules}

An additional effect that required modelling in our NLTE spectra compared to the models in \citet{Barmentloo_2024_Nitrogen} is the \ajc{formation and cooling} of molecules in the SN nebula. The molecule CO has been identified in a large number of Type II SNe (e.g. \citet{Spyromilio_1996_COovertone, Gerardy_2000_COovertone, Spyromilio_2001_COovertone}) and Type I SNe (e.g. \citet{Gerardy_2002_COovertoneI, Hunter_2009_COovertoneI, Banerjee_2018_COovertoneI}). Models (e.g. by \citet{Liu_1995_Molecules}) have shown that CO formation can cause a noticeable \ajc{enhancement} of the cooling of the ejecta in the O/C zone, which is one of the two main zones where carbon emission originates (see Section \ref{sec:results_line_formation}). While a simplified treatment of \ajc{molecular effects} was present in \citet{Jerkstrand_2015_Canon}, the detailed creation and emission of molecules was added to \SUMOcodename{} by \citet{Liljegren_2020_MoleculesI} and its effects on nebular SESNe were studied in \citet{Liljegren_2023_MoleculesinSESNe}. In our models, molecule formation and cooling was treated as described in those works. The effects of molecule formation on carbon emission are analysed in Section \ref{sec:results_molecules}.

\begin{figure*}
    \centering
    \includegraphics[width=0.475\linewidth]{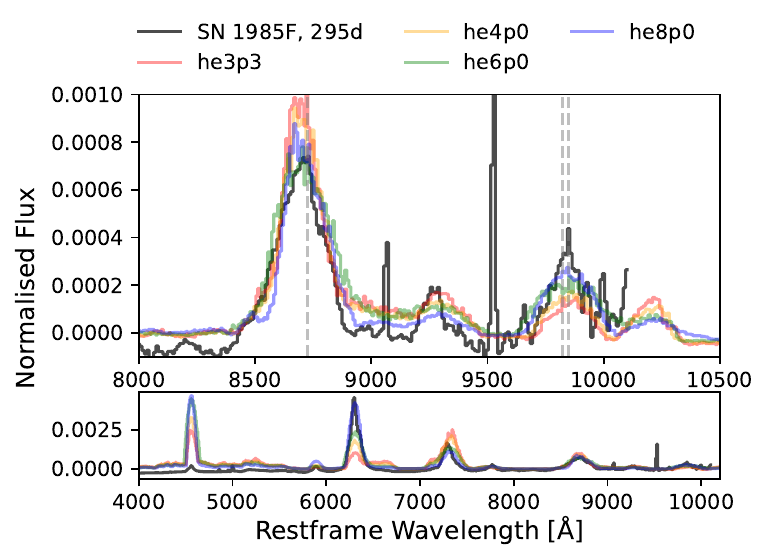}
    \includegraphics[width=0.475\linewidth]{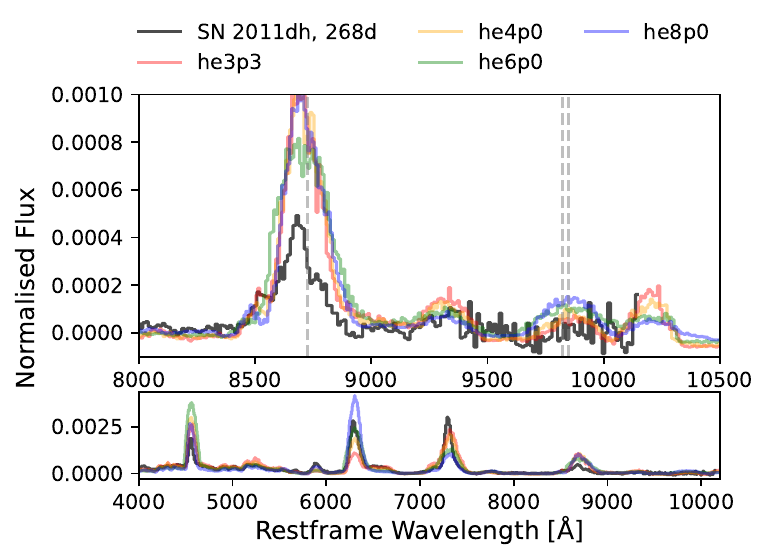}\\[1ex]
    \includegraphics[width=0.475\linewidth]{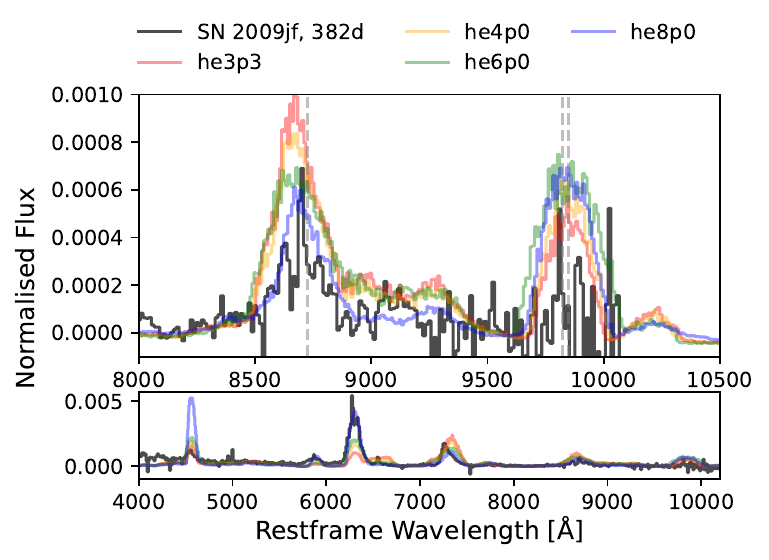}
    \caption{Three examples of spectra in our NIR sample, compared to synthetic spectra. In each example, the top subplot shows a zoom in of the emission region containing \CIsinglet{} and \CIdoublet{} (both indicated by vertical dashed lines), while the bottom subplot showcases the range from 4 000 to 10 200 Å. Spectra have been normalised to the optical luminosity (see Equation \ref{eq:definition_9850_diagnostic}). In the first subplot, \CIdoublet{} is detected and both lines have strengths similar to the models. In the latter two subplots, \CIdoublet{} is not detected, and \CIsinglet{} in the \CaNIRtriplet{} complex is clearly weaker than in the model spectra. }
    \label{fig:three_examples}
\end{figure*}

\subsubsection{Filling Factors}
\label{sec:filling_factors}

The heating of the radioactive decay of \Ni{} and \el{56}{Co} causes the inner Fe/He and Si/S zones to expand during the first days after explosion, which is known as the "Ni-bubble effect" \citep{Herant_1991_Nibubble,Gabler2021}. This leads to a density contrast between these zones and the remaining macroscopically mixed zones. As this effect is not present in the ejecta models, it needs to be parameterised when preparing our ejecta models for \SUMOcodename{}. Estimates of the resulting densities (or filling factors, i.e. the fraction of the core-volume that is occupied by a specific core zone) have been made for individual SNe \citep[e.g. for SN 2004et,][]{Jerkstrand_2012_SUMOb}, but a systematic study of filling factors in SNe is still lacking. Filling factors should thus be treated as an unknown parameter.

The effect of filling factors on \LCIdoublet{} and \LCIsinglet{} was studied for three different model sets. These had density contrasts 1-5-20-20-20-20 (model $\chi = 20$), 1-10-30-30-30-30 (model $\chi = 30$) and 1-10-100-100-100-100 (model $\chi = 100$). Here, the density contrast is defined as the overdensity compared to the Fe/He zone. So, in model $\chi = 30$, the density in the Si/S zone is ten times that of the density in the Fe/He zone, and for the O/Si/S, O/Ne/Mg, O/C zones and the mixed-in part of the He/C zone, the density is thirty times that of the Fe/He zone. \updated{The closing constraints are that the total mass in the macroscopically mixed core is unchanged, just as the mass of each individual zone in the core and the total volume of the core. With this in mind, the equation to determine the filling factor ff$_{i}$ for a core zone \textit{i} with mass $M_{i}$ becomes:}

\begin{equation}
    \text{ff}_{i} = \frac{ \frac{M_{i}}{\chi_{i}}    }{\sum^{n}_{i = 1} \frac{M_{i}}{\chi_{i}}}
\end{equation} 

\updated{With zone $i = 1$ the Fe/He zone, and $n$ the number of core zones. The change in densities that the filling factor approach leads to in the ejecta models is given in the tables in Appendix \ref{app:compmodels} as the clumping factor, f$_{\text{clumping}}$.} Of the three sets that we studied, model $\chi = 30$ will be used as our standard set throughout this work (following \citet{Jerkstrand_2015_Canon} and \citet{Barmentloo_2024_Nitrogen}). The effects of filling factors on carbon emission are analysed in Section \ref{sec:results_fillingfactor}.

\subsubsection{Carbon Atomic Data}

The model atom used in \SUMOcodename{} to simulate carbon contains the first 91 energy levels and has a total of 793 transitions. Of these, 17 levels have specific recombination rates \citep{Nahar_1995_RRCI} and 10 transitions have specific collision strengths\footnote{Obtained through \url{https://www.astronomy.ohio-state.edu/pradhan.1/}} (both exist for \CIdoublet{} and \CIsinglet{}). The total radiative recombination rate from C II to C I is again taken from \citet{Nahar_1995_RRCI}. For most photoionisation (PI) cross sections (\updated{which we find to dominate over non-thermal and collisional ionisation for C I in our models}), we use the hydrogenic approximation. The exceptions are the ground state cross section, which is based on the \citet{Verner_1996_Photoionisation} routine, and the $^{1}$S and $^{5}S°$ states, which are taken from TOPBASE\footnote{\url{https://cdsweb.u-strasbg.fr/topbase/xsections.html}}.<\ajc{We carried out some sensitivity tests indicating that reasonable variation of the hydrogenic cross sections led to differences in carbon line luminosities} of $\lesssim$ 5$\%$. The transition probabilities for \CIsinglet{} ($A=6\times 10^{-1}$ s$^{-1}$) and \CIdoublet{} ($A=7.3\times 10^{-5}$ s$^{-1}$ and 2.2 $\times$ 10$^{-4}$ s$^{-1}$, for the respective components) are provided by \citet{Fischer_2006_CI_TP_data}. The uncertainty that NIST assigns to the transition probabilities is $\le$ 10\% for \CIsinglet{} and $\le$ 18\% for the components of \CIdoublet{}.

\updated{In this work, we do not consider charge transfer in our simulations. Previous modeling has found the C I + O II  $\rightarrow$ C II + O I reaction to be important in SN ejecta \citep{Jerkstrand_2011_ct, Jerkstrand_2015_Canon}. However, in that work the authors use a heavily simplified estimate for the reaction rate. Detailed quantum mechanical calculations were recently performed (Reja et al. 2025, in prep.) for the rates of the C I + O II  $\rightarrow$ C II + O I reaction, and it was found that the reaction rate is significantly lower than the estimate in \citet{Jerkstrand_2011_ct}. Nonetheless, including the reaction with the quantum mechanical rates is found to lead to a decrease in carbon luminosities of $\lesssim$ 10\% compared to when it is not included \updated{for the phases considered here}, mostly due to a decreased neutral carbon fraction in the O/C zone. }

\subsection{Estimating \LCIdoublet}
\label{subsec:estimating_9850}

Estimating \LCIdoublet{} should in principle be rather trivial. As seen in an ion-by-ion decomposition of the model spectra for the he4p0 model\footnote{The general picture is the same for all our other models} in Figure \ref{fig:ion_by_ion}, \CIdoublet{} is found to be an effectively isolated line in SESNe. The only exceptions to this are early epochs in low mass models, where \ion{Co}{II} can contribute significantly to the red wing of \CIdoublet{}. However, the blue half of the doublet alone should still allow for a reasonable Gaussian fit (see Appendix \ref{app:doublet_is_observed} and Figure \ref{fig:validating_9850}).

When setting out to measure \LCIdoublet{}, we noticed that a clear emission line at the location of \CIdoublet{} was missing in the majority of our observed spectra (see two examples in Figure \ref{fig:three_examples}). In some cases this was due to poor S/N in the spectra, but even for those with good S/N there was often no sign of \CIdoublet{}. 

The lack of \CIdoublet{} only allowed us to set upper limits on \LCIdoublet{} for most objects.\footnote{In the analytical formalism described in \ref{sec:theory}, this means that for most objects we can effectively only obtain an upper limit on \McarbonI{}.} We determined the upper limits by following the methodology outlined in \citet{Shappee_2013_UpperLimit} and \citet{Lundqvist_2015_UpperLimit}. Our exact methodology is described in detail in Appendix \ref{app:doublet_luminosity}.

\subsection{Estimating \LCIsinglet}
\label{subsec:estimating_8727}

The spectral region where \CIsinglet{} is located is filled with at least six well known emission lines (see Table \ref{table:six_lines}). This fact makes inferring \LCIsinglet{} \ajc{challenging} \citep{Fransson_1989_Canon, Mazzali_2010_2007grcarbon, Jerkstrand_2015_Canon}. In theory, one could use \SUMOcodename{} to obtain an estimate for \LCIsinglet{}: by creating a sufficiently large grid of models with different carbon abundances and comparing each model \CaNIRtriplet{} complex line profile to the observed one, one could find the input model giving the best fit and obtain how much of the \CaNIRtriplet{} complex luminosity was caused by \CIsinglet{}. \ajc{This approach however becomes limited by inaccuracies in the Ca II modeling, in addition to C I. It is also} very computationally expensive, with a single \SUMOcodename{} epoch requiring on the order of 100 CPU hours to converge. Creating a grid with sufficient free parameters (as well as covering nebular epochs with sufficiently small intervals) thus becomes intractable. Furthermore, if a reader would want to use this method themselves but their desired parameter combination is not present in the initial grid, extending the grid would be cumbersome.

In this work we present a novel approach to estimating \LCIsinglet{}. We created \ajc{a new, simple and fast, Monte Carlo transfer code},
which we dub \codename{}\footnote{\codename{} will be made publicly available for anyone to use at \url{https://github.com/StanBarmentloo/CaNARY}}. This code solely simulates the physics relevant for the \CaNIRtriplet{} region, reducing the CPU time for creating a single \CaNIRtriplet{} complex line profile from 100 CPU hours (as part of a full \SUMOcodename{} model) to a mere 30 seconds. 

In short, the code simulates the ejecta as an expanding sphere of material with a single outer-velocity. The material in the sphere is assumed to consist only of the elements mentioned in Table \ref{table:six_lines}, and these elements can either emit photons (at a given luminosity) or scatter them (with a given optical depth), or both. The resulting simulation has a total of six free parameters (significantly fewer than fitting six Gaussians) and still includes much of the relevant physics to the problem. For this work, we used the code to create a spectral library of $\sim$ 3000 \CaNIRtriplet{} complex line profiles. Subsets of these were compared to each observed spectrum by determining a $\chi^{2}$-score. By finding the minimum in the $\chi^{2}$ vs. \LCIsinglet{} curve, an estimate for \LCIsinglet{} was then determined. A more detailed description of \codename{} is given in Appendix \ref{app:singlet_luminosity}.

\section{Results}
\label{sec:results}

\subsection{Carbon line formation in SESNe}
\label{sec:results_line_formation}

\subsubsection{General description of carbon line formation in the models}
\label{sec:general_formation}

\begin{figure}
    \centering
    \includegraphics[width=0.475\textwidth]{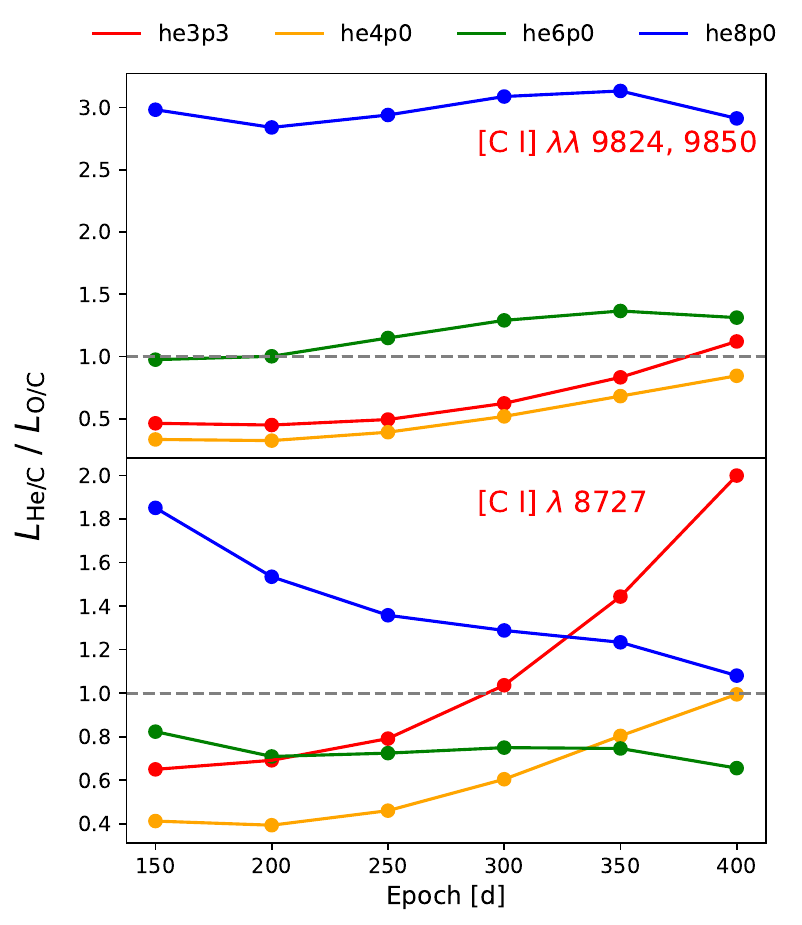}
    \caption{The ratio of the luminosity of both carbon lines in the He/C and O/C zones for our models. The relative importance of the He/C zone is found to increase with core mass, and for the lower mass models also with time. }
    \label{fig:hec_vs_oc_flux_ratio}
\end{figure}

\begin{figure}
    \centering
    \includegraphics[width=0.475\textwidth]{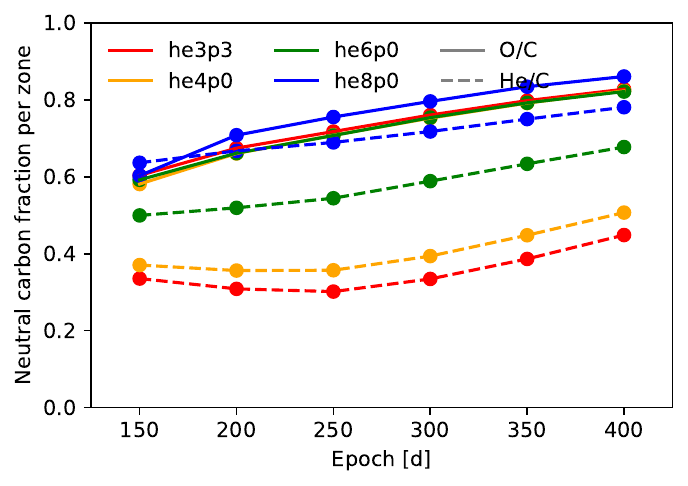}
    \caption{The \ion{C}{I} fraction for the He/C and O/C zones in our models. While the neutral fraction is found to evolve roughly the same for all helium core masses in the O/C zone, it is consistently higher for increasing helium core mass in the He/C zone. }
    \label{fig:hec_vs_oc_ionisation}
\end{figure}

For all models in our grid, the majority ($\gtrsim$ 90\%) of \CIsinglet{} and \CIdoublet{} emission originates from two zones: O/C and He/C. Whereas the O/C zone is typically more dense (\updated{by roughly an order of magnitude}) and has a higher neutral carbon fraction (see Figure \ref{fig:hec_vs_oc_ionisation}), the He/C zone is typically hotter. The resulting ratio of He/C to O/C carbon luminosity is shown in Figure \ref{fig:hec_vs_oc_flux_ratio}. 

\updated{For \CIdoublet{}, we see that with increasing core mass, the importance of the He/C vs O/C zone increases. We attribute this to a combination of increased relative mass of the He/C zone \ajc{(see Table \ref{tab:main_properties})} and an increased neutral carbon fraction in the He/C zone. For \CIsinglet{}, the picture is not as clear, which reflects that other parameters, such as the zone's temperature and density, dominate in setting the luminosity ratio of these two zones. Lastly, for both lines the lower mass models show an increase of the He/C to O/C luminosity ratio with time. }

\subsubsection{Molecules}
\label{sec:results_molecules}

\begin{figure}
    \centering
    \includegraphics[width=0.475\textwidth]{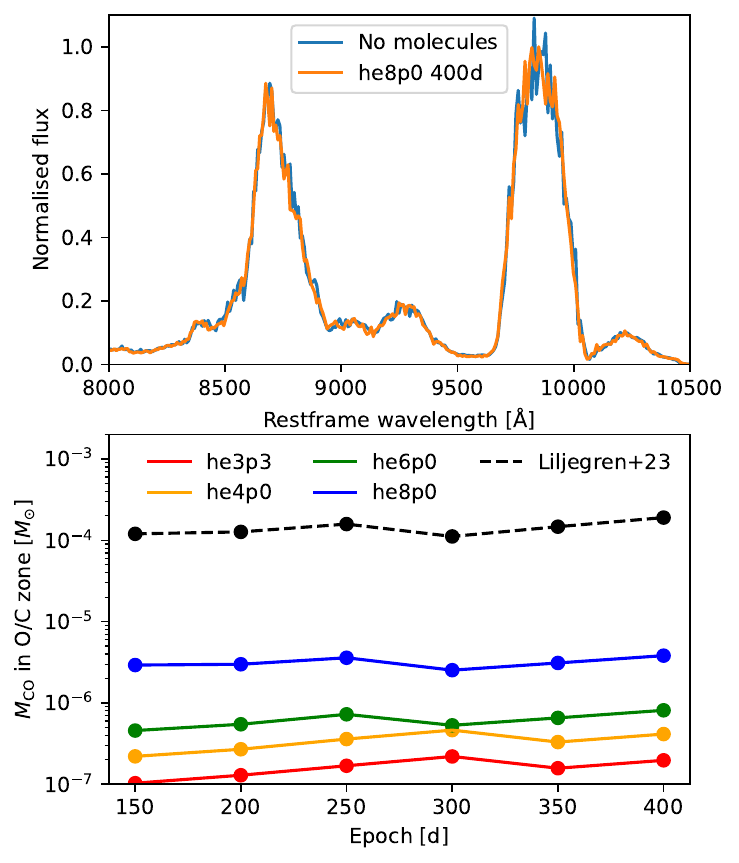}
    \caption{\textit{Top panel:} Comparison of the standard ($\chi$ = 30) he8p0 model at 400d with and without molecules turned on. The effect of molecule formation is negligible for \CIsinglet{} and \CIdoublet{} in our model grid. \textit{Bottom panel:} Masses of CO formed in the O/C-zone in our model grid. The black dotted line indicates the values for the Type Ic model in \citet{Liljegren_2023_MoleculesinSESNe} (see Section \ref{sec:molecules}).}
    \label{fig:molecule_formation}
\end{figure}

In the top panel of Figure \ref{fig:molecule_formation}, a comparison of spectra with- and without molecule formation switched on is presented for the he8p0 model (which is the model with the highest molecule masses) at 400d. The figure makes it clear that molecules do not strongly affect the luminosity of carbon lines in our model grid. This can be attributed to the small molecule production in our models (bottom panel of Figure \ref{fig:molecule_formation}); comparing the CO masses in our most molecule-rich model (he8p0) to the Type Ic model in \citet{Liljegren_2023_MoleculesinSESNe}, we obtain $\sim$ 40 -- 50 times less CO in the O/C zone. 
A smaller O/C zone mass (0.36 vs 1.4 \Msun{}) combined with a lower density in this zone (3.2 $\times$ 10$^{8}$ vs 1.1 $\times$ 10$^{9}$ cm$^{-3}$ at 200d, leding to less efficient CO production) explains much of this difference.

It should be noted that even for the \citet{Liljegren_2023_MoleculesinSESNe} model (of which the first CO overtone emission compares well to observations; see their figure 17), the largest temperature difference caused by molecules for the epochs studied in our work ($\sim$ 300 K, see their figure 4) would cause a $\sim$ 30\% decrease of \CIsinglet{} emission \textit{in the O/C-zone} (and $\sim$ 15\% for \CIdoublet{}). For our model with the lowest He/C to O/C luminosity ratio the effect on the total carbon luminosities would be $\lesssim$ 20\% ($\sim$ 10\% for \CIdoublet{}).

\subsubsection{Filling factor}
\label{sec:results_fillingfactor}

\begin{figure}
    \centering
    \includegraphics[width=0.475\textwidth]{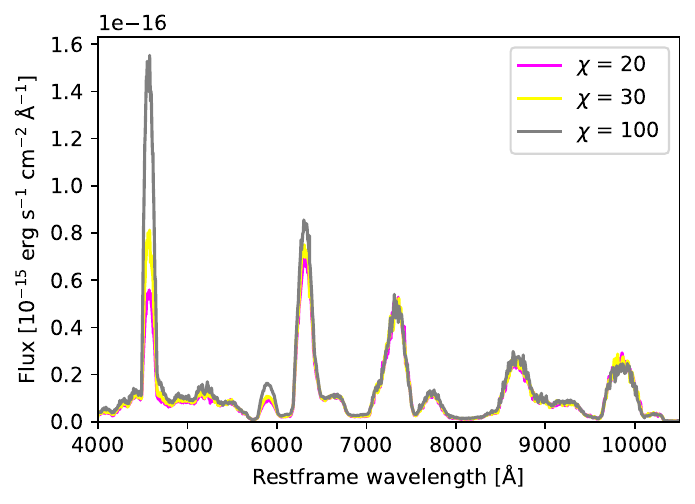}
    \caption{A comparison of the same input model (he6p0 at 400d) for the three different filling factors described in Section \ref{sec:filling_factors}. As described in Section \ref{sec:results_fillingfactor}, the figure shows that while uncertainty in filling factors does have its influence on some emission lines, it does not greatly affect the model carbon luminosities.}
    \label{fig:filling_factors}
\end{figure}

A comparison for our three different filling factor setups (see Section \ref{sec:filling_factors}) for he6p0 at 400d (the model and epoch for which different filling factors led to the largest spectral differences) is shown in Figure \ref{fig:filling_factors}.
As the figure shows, clear differences arise in the emission strengths of the prominent Na I D and \ion{Mg}{I}] $\lambda$ 4571 emission lines. For these lines, this originates mostly from the fact that for larger $\chi$, the neutral fractions of these elements increase considerably. However, uncertainty in filling factors has no significant effect on carbon emission. For \CIdoublet{} this statement is most robust, as its upper level is always close to LTE\footnote{\updated{Strictly speaking one should differentiate between LTE w.r.t. the whole atom and LTE w.r.t the ground state, but for our applications here this difference is $\lesssim$ 1\%, which is less than the number of significant digits we quote.}} (departure coefficient $>$ 0.95) for the models in our grid. Different densities then lead to two counteracting effects on the upper level population; higher $\chi$ leads to lower temperatures (decreasing the population) and to a higher neutral carbon fraction (increasing the population). From the figure it is clear that these effects quite closely balance each other in our grid. It should be added that filling factors only influence the carbon emission from the O/C zone (and the mixed-in part of the He/C-zone). In conclusion, we find that the uncertainty in filling factors for the carbon-rich zones does not significantly affect our inferences of \LCIsinglet{} and \LCIdoublet{}.

\subsection{\LCIdoublet}
\label{sec:results_9850}

We performed our methodology for estimating (upper limits on) \LCIdoublet{} described in Section \ref{subsec:estimating_9850} and Appendix \ref{app:doublet_luminosity} for all spectra in the NIR sample. To be able to compare the absolute luminosities across the range of ejecta masses that these SNe cover, we need a diagnostic that normalises for the total luminosity in the spectra. Similar as to what was done in \citet{Dessart_2021_SESNespectra}, \citet{Barmentloo_2024_Nitrogen} and \citet{Fang_2025_RSGproblem}, we do this by dividing by an 'optical luminosity', as in Equation \ref{eq:definition_9850_diagnostic}.
\begin{equation}
    f_{[\ion{C}{I}]\lambda\lambda9824,9850} = \frac{\int F_{[\ion{C}{I}]\lambda\lambda 9824,9850}} {\int^{8000\  \text{Å} }_{5000\ \text{Å}} (F_{\lambda} - F_{\text{pseudo}}) d\lambda} \times 100
    \label{eq:definition_9850_diagnostic},        
\end{equation}
where $F_{[\ion{C}{I}]\lambda\lambda 9824,9850}$ is the best fit (upper limit-) gaussian for the \CIdoublet{} line, $F_{\lambda}$ is the observed flux in the spectrum and $F_{\text{pseudo}}$ is the pseudo-continuum. This pseudo-continuum is defined as the average of the three lowest mean fluxes found in the regions 5740 -- 5790 Å, 6020 -- 6070 Å, 6850 -- 6900 Å and 7950 -- 8000 Å (see appendix A3 in \citet{Barmentloo_2024_Nitrogen}). \updated{The reason for removing this pseudo-continuum is that in most observed spectra, there will be an unknown amount of background emission from the host galaxy. Naturally, the model spectra will not have this additional emission. For most spectra the host contribution is low, but to assure an as equal as possible comparison between observations and models, we remove the pseudo continuum for all spectra. Comparing \CIdoubletdiag{} results with the pseudo continuum subtraction turned on and off (not shown here), we find that none of our general conclusions in this work are altered. }

\begin{figure*}
    \centering
    \includegraphics[width=0.95\linewidth]{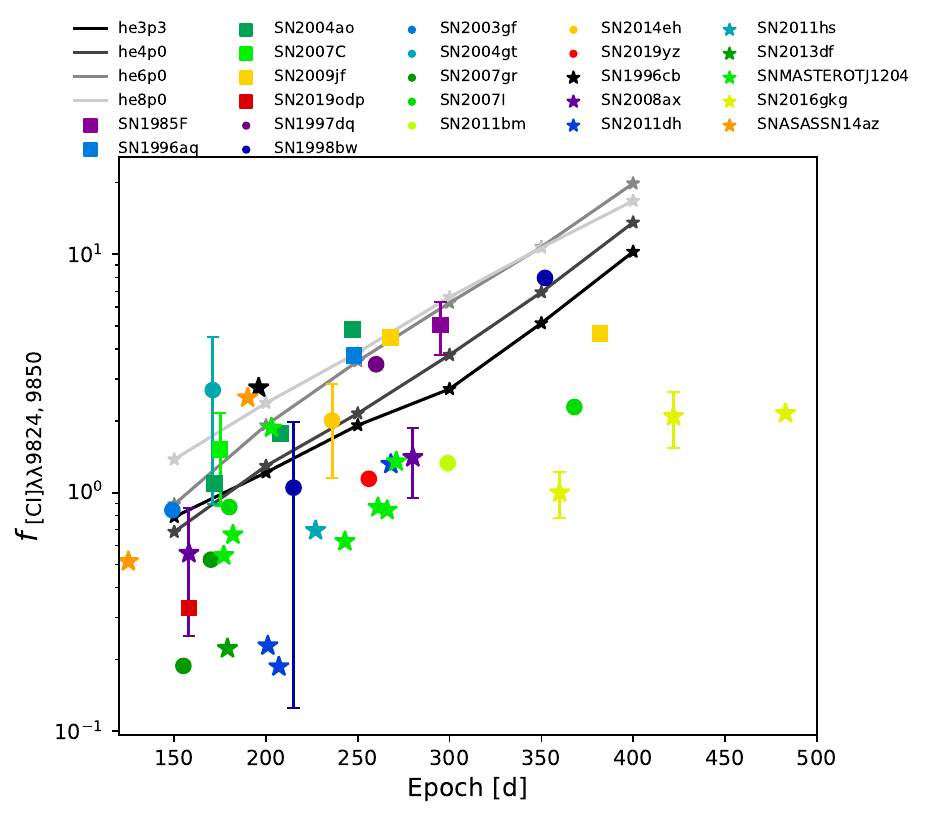}
    \caption{The \CIdoubletdiag{} estimates as function of time for both our model set and our SESNe sample, obtained using the methodology described in Appendix \ref{app:doublet_luminosity}. Squares indicate Type Ib, circles indicate Type Ic and stars indicate Type IIb SNe. Data points with error bars indicate that \CIdoublet{} was detected; those without error bars are thus upper limits. To avoid cluttering, only the data points of the model spectra were connected with lines. For more than half of the observed spectra, we estimate \CIdoubletdiag{} values below our lowest model curve. Note the logarithmic y-axis.}
    \label{fig:CI9850_results}
\end{figure*}

Figure \ref{fig:CI9850_results} presents the computed \CIdoubletdiag{} values as a function of epoch for each observed spectrum, as well as for our synthetic spectra. 

Considering the plot as a whole, the first impression is that the values for \CIdoubletdiag{} in observed spectra tend to average lower than in synthetic spectra. In fact, many of the observations lie below even the lowest model curve, with many of these in turn being merely upper limits. This might have been expected from our earlier observation in Section \ref{subsec:estimating_9850}, where in very few cases \CIdoublet{} was actually observed (9 out of 47). Of those nine detections (all shown in Figures \ref{fig:hec_sizea} and \ref{fig:hec_sizeb}), five are found to lie below the model curves, three are within the studied range, and one (SN 2004gt) is above all of the model curves. 

\updated{What also becomes clear is that in the model spectra the emission of \CIdoublet{} as part of the full spectrum increases exponentially with time, with higher mass models having higher \CIdoubletdiag{} values for a given epoch. Although we do not have any SNe with a sufficient amount of detections to confirm this exponential trend, the nine detections that we do have are consistent with an increase of \CIdoubletdiag{} with time.} These findings make late ($\gtrsim$ 300d) epochs the most effective at determining accurate carbon abundances in SNe. Unfortunately, the continuously decreasing brightness of SNe means that observed spectra become increasingly scarce here. Of the eight spectra in our sample that have epochs $\gtrsim$ 300d, we find six to have \CIdoubletdiag{} values below the lowest model curve, with only a single detection (SN 1985F) within the model range. 

\subsection{\LCIsinglet}
\label{sec:results_8727}

\begin{figure*}
    \centering
    \includegraphics[width=0.99\textwidth]{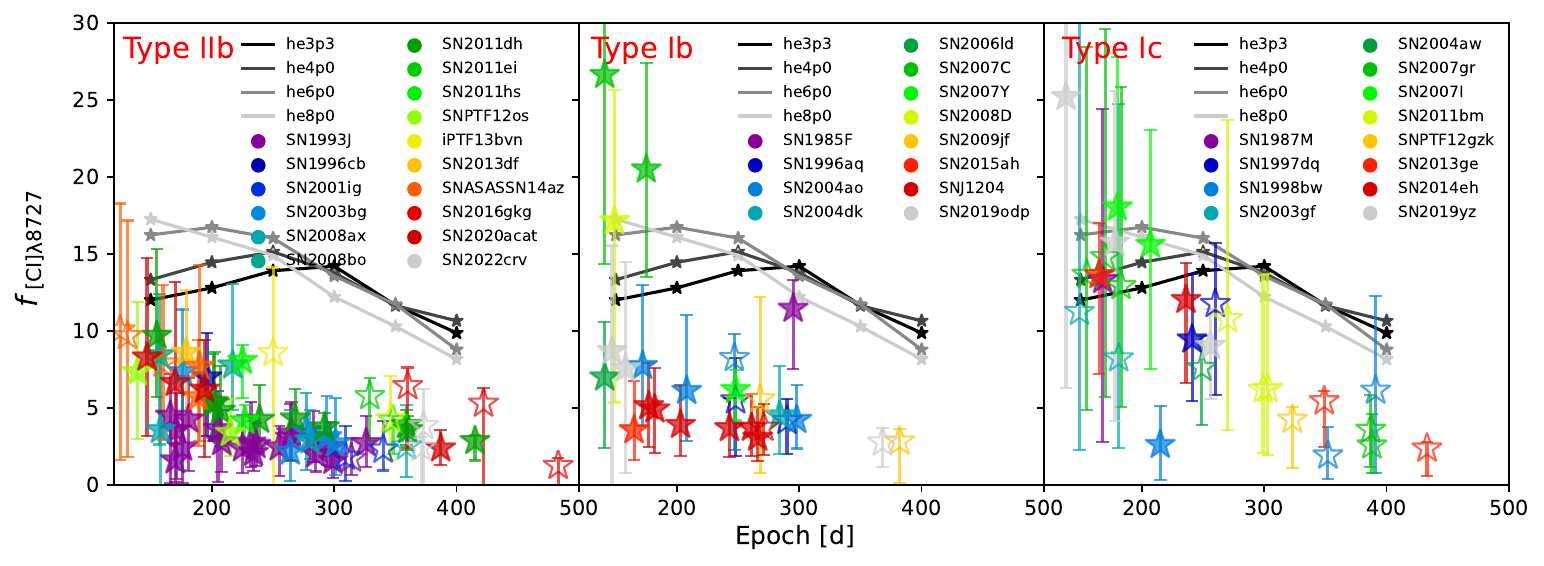}
    \caption{The \CIsingletdiag{} estimates obtained by \codename{} for the SNe in our sample, as well as for our model spectra. Empty stars indicate spectra where the \CaNIRtriplet{} region was deemed particularly noisy (RMSE$^{2}$ $\gtrsim$ 0.015; see the text in Section \ref{sec:results_8727}). The error bars are calculated as described in Section \ref{app:model_selection}  To avoid cluttering, the SNe have been separated into their three subtypes. As was also noted for \CIdoubletdiag{}, many of the \CIsinglet{} estimates for the SNe in our sample are found to lie outside of the model tracks, especially for the Type IIb and Type Ib SNe.}
    \label{fig:CI8727_results}
\end{figure*}

Analogously to the previous section, Figure \ref{fig:CI8727_results} shows the temporal evolution of \CIsingletdiag{}, which follows \ajc{an analogous} definition as Equation \ref{eq:definition_9850_diagnostic}. The figure has been split into three (one panel for each SESN subtype) to avoid cluttering due to the increased number of datapoints. During the process of estimating \LCIsinglet{}, it was found that a non-negligible amount of spectra were of poor quality around 8727 Å, which made interpreting the fits produced by \codename{} more difficult. Such objects are indicated as empty stars and are defined as having RMSE$^{2}$ $\gtrsim$ 0.015. 

A clear trend arises when moving through the sequence IIb $\rightarrow$ Ib $\rightarrow$ Ic, namely that the mean \CIsingletdiag{} value becomes increasingly larger. For Type IIb SNe, none of the 67 spectra have a \CIsingletdiag{} value within the range of model curves. For Type Ib SNe the picture is mostly the same, but now 4 out of 28 are around or above the synthetic spectra tracks. Finally the Type Ic SNe sample (which has the highest percentage of RMSE$^{2}$ $\gtrsim$ 0.015 spectra) is mostly compatible with the modeled curves. This trend is of particular interest, as Type Ic progenitors are typically thought of as lacking a helium zone at the time of explosion. We discuss this point further in Section \ref{subsec:hec_size}. Compared to \CIdoublet{}, the model tracks do not show any strong correlation with helium core mass or epoch. \updated{It should further be noted that the luminosity prediction for \CIdoublet{} is less sensitive to modeling details, as its lower upper state energy (1.26 eV vs. 2.68 eV) and lower critical density (2.6 $\times 10^{4}$ vs. 2.5 $\times 10^7$ g cm$^{-3}$) makes the population of the upper state less sensitive to temperature and non-LTE effects.} All in all, just as for \CIdoublet{}, we find relative luminosities of carbon lines in observed SNe to be lower than for the model spectra produced here.

\subsection{\Mcarbon{} Estimates}
\label{subsec:carbon_mass_estimation}

\begin{figure*}
    \centering
    \includegraphics[width=0.975\linewidth]{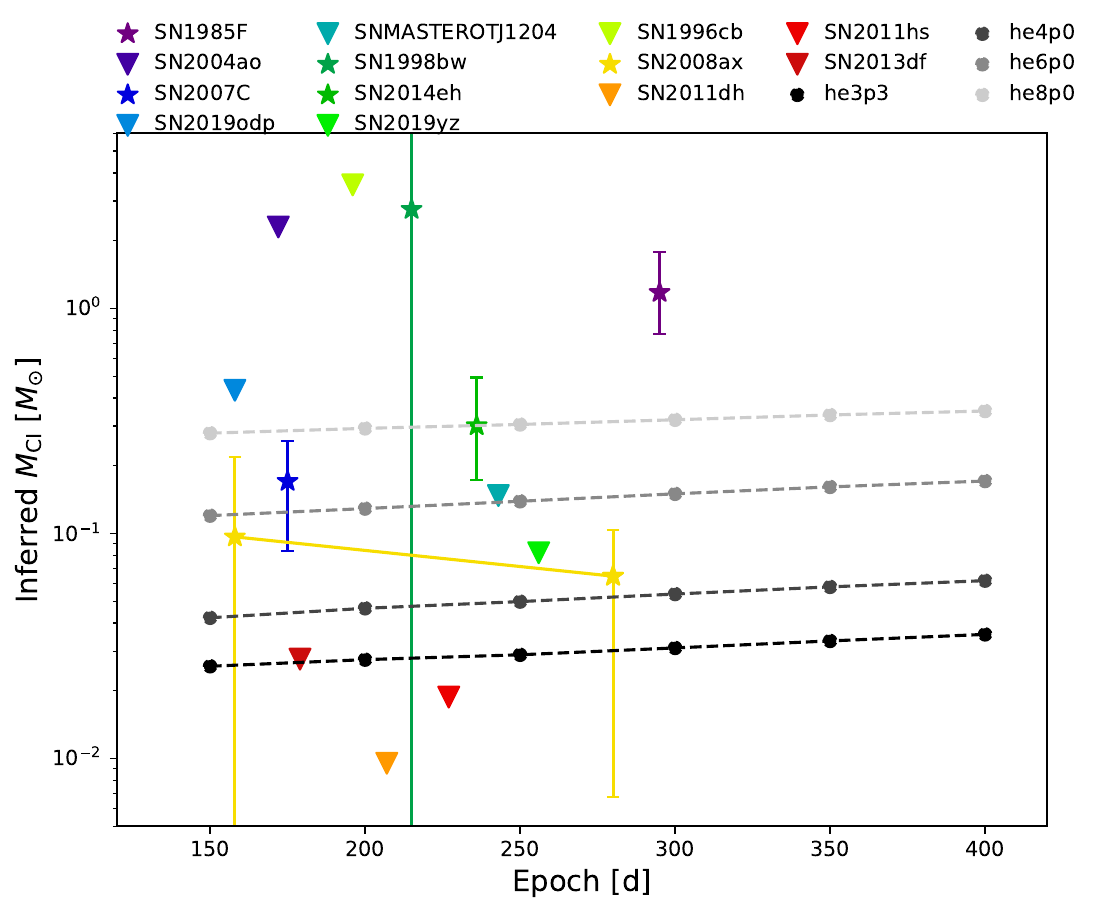}
    \caption{Estimates of \McarbonI{} for the SNe in our sample, compared to the actual masses in our model grid. Note the logarithmic y-axis. Upper limits (caused by an upper limit on \CIdoublet{}) are indicated by downward facing triangles. The error bars include uncertainties in \LCIsinglet{}, \LCIdoublet{} and the $A$ values for both lines, added using standard error propagation. The error bars do not include uncertainties in host extinction, distance, or photometry. Different from what was found in Figures \ref{fig:CI9850_results} and \ref{fig:CI8727_results}, the mass estimates evenly cover the entire range present in the models.}
    \label{fig:carbon_mass_estimates}
\end{figure*}

\begin{figure}
    \centering
    \includegraphics[width=0.975\linewidth]{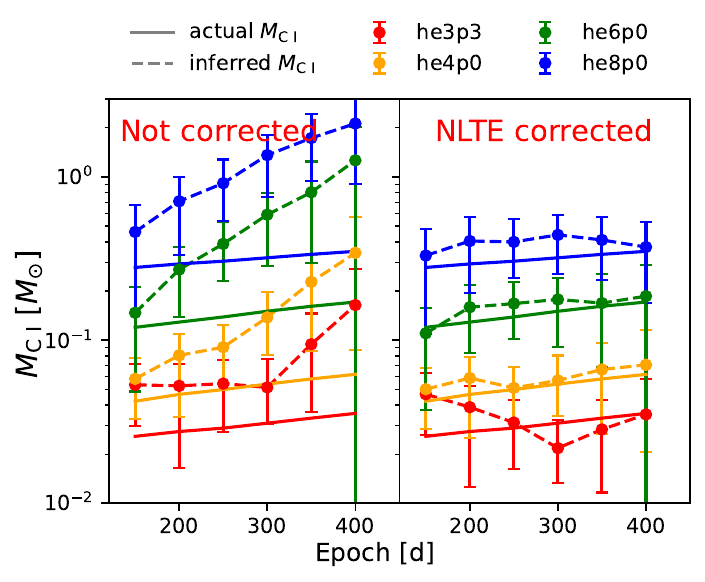}
    \caption{\textit{Left panel:} The actual \McarbonI{} masses in our models compared to estimates of \McarbonI{} obtained using the analytical, LTE formalism described in Section \ref{sec:theory}. Masses are overestimated increasingly with time as the parent state of \CIsinglet{} falls out of LTE. \textit{Right panel:} Same as left panel, but now with our estimates corrected for the departure from LTE of \CIsinglet{} as presented in Figure \ref{fig:lte_departure}. With this correction applied, our estimates differ from the actual values by at most $\sim$ 30\%, showcasing that all assumptions made in the methodologies of estimating \LCIsinglet{} and \LCIdoublet{} do not lead to too large errors. The error bars include uncertainties in \LCIsinglet{} and the A values for both lines.   }
    \label{fig:007_Model_Mass_Estimate_Validation}
\end{figure}

\begin{figure}
    \centering
    \includegraphics[width=0.85\linewidth]{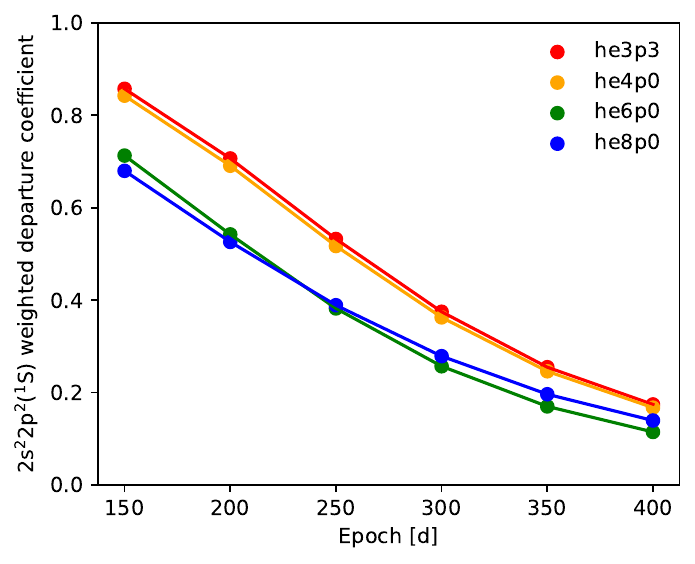}
    \caption{Departure coefficients for the parent state of \CIsinglet{} in the O/C and He/C zones in our models. The coefficients are weighted by the \McarbonI{} mass in each zone. Higher mass helium cores have stronger departures, as their He/C zones are relatively more massive (see also Figure \ref{fig:hec_vs_oc_flux_ratio}).}
    \label{fig:lte_departure}
\end{figure}

As discussed in Section \ref{sec:theory}, we can combine the values for \LCIsinglet{} and \LCIdoublet{} to arrive at an estimate of the neutral carbon mass. To test this methodology, we first made these estimates for our model spectra, for which we know the true neutral carbon mass present in the ejecta. The resulting estimates are shown in the left panel of Figure \ref{fig:007_Model_Mass_Estimate_Validation}. It is clear that applying the analytical formalism unchanged leads to neutral carbon mass estimates which are about $\sim$ 2 to 10 times too high. The clue for the cause comes from the upward trend in time; we know from our models (see Figure \ref{fig:hec_vs_oc_ionisation}) that total ionisation fractions do not change appreciably, and should therefore obtain roughly the same mass estimate at each epoch. It turns out that (as alluded to in Section \ref{sec:theory}) the assumption that the upper level of \CIsinglet{} is in LTE does not hold. In fact, it does not hold even for our earliest epoch, with departure coefficients at $\sim$ 0.7 -- 0.9 (see Figure \ref{fig:lte_departure}). 
\updated{This means that the luminosity of \CIsinglet{} is no longer described by Equation \ref{eq:lte_luminosity}, but rather by Equation \ref{eq:nlte_luminosity}, which is dependent on the electron density $n_{e}$ in the relevant zone. } Now, for our model spectra we can use this knowledge of the departure coefficient to 'correct' the upper level population going into Equation \ref{eq:lte_luminosity}. When one does so, one obtains neutral carbon mass estimates as shown in the right panel of Figure \ref{fig:007_Model_Mass_Estimate_Validation}. The improvement is drastic, with typical discrepancy down to $\lesssim$ 30\%. From this one can conclude that the effect of the remaining assumptions (all emission originating from a single zone with a single temperature, as well as all simplifications going into the methodology for estimating \LCIsinglet{} and \LCIdoublet{}) on the mass estimate is tractable. 

Now we turn to our observations. To attempt to remove the LTE departure effect, we 'correct' the upper level populations for \CIsinglet{} in the same way as for our models. Assuming the departure coefficient to be the same for the ejecta in our observations is a large assumption at best and we can therefore not stress enough that the obtained neutral carbon masses should be taken with a grain of salt. That said, the departure coefficient evolves in a similar way in all our models. Furthermore, as most of our mass estimates will be upper limits (because of the many non-detections of \LCIdoublet{}), we opt to correct the upper level for our entire sample using the model departure coefficient track that leads to the most conservative upper limits. This is the track of the he3p3 model. Besides this correction, two sources of uncertainty enter that were not present when comparing the models amongst themselves, namely extinction of and distance to the SN. The extinction estimates for some of our objects (see Section \ref{subsec:extinction} and Table \ref{tab:NIR_sample}) are poorly constrained, in which cases we only considered extinction by the MW. For these objects, we may thus be underestimating the estimated mass significantly. Keeping all of these factors in mind, the neutral carbon mass estimates for our SESN sample are presented in Figure \ref{fig:carbon_mass_estimates}.

The picture is different than found in Figures \ref{fig:CI9850_results} \& \ref{fig:CI8727_results}: the sample now covers the parameter space more evenly. That said, there are still three SNe (all of Type IIb; SN 2011dh, SN 2011hs and SN 2013df) with upper limits on the estimated mass below even our lowest mass model. At the same time, for two SNe with detected \CIdoublet{} emission we estimate high neutral carbon masses of 1.2 (Type Ib SN 1985F) and 2.6 \Msun{} (Type Ic-BL SN 1998bw). In fact, both of these SNe are thought to originate from massive progenitors with $M_\mathrm{C} \sim$ 1 \Msun{} \citep{Fransson_1989_Canon, Woosley_1999_1998bw, Patat_2001_1998bw}. However, our estimates are for \McarbonI{}, so with typical neutral carbon fractions of $\sim$ 0.5 -- 0.7, they would still be too high.

\begin{figure}
    \centering
    \includegraphics[width=0.975\linewidth]{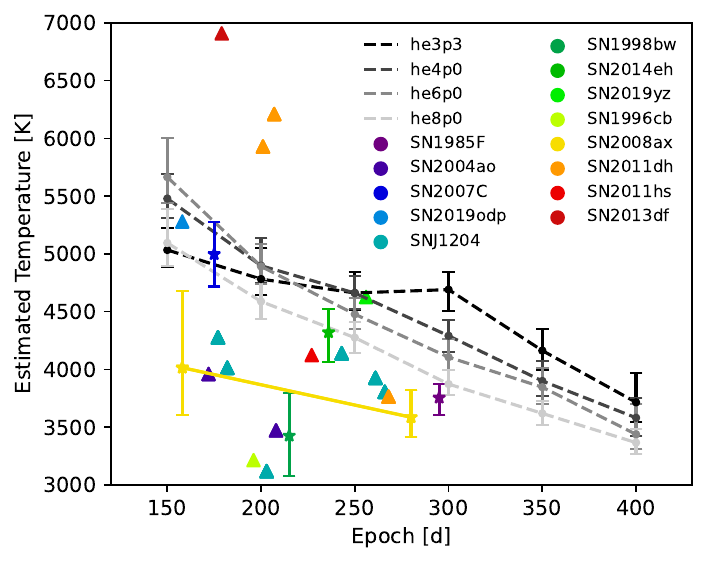}
    \caption{Temperature estimates for the carbon in our observed and model spectra, obtained through the methodology described in Section \ref{sec:theory}. Upward pointing triangles indicate lower limits. All temperature estimates in the plot have been corrected for the NLTE departure of \CIsinglet{}, as described in Section \ref{subsec:carbon_mass_estimation}. The error bars include the same sources of uncertainty as in Figure \ref{fig:carbon_mass_estimates}.}
    \label{fig:temperature_estimates}
\end{figure}

It is also of interest to take a look at the estimated temperatures that lead to the mass estimates. These are shown in Figure \ref{fig:temperature_estimates}, and have all again been corrected for the LTE departure. For most SNe in the sample, our estimated temperatures are significantly lower than in our models, even when only considering those with \CIdoublet{} detections. One could argue that the \SUMOcodename{} models cool too slowly (through e.g. too little molecular cooling), and that this is the sole reason for the higher relative emission of \CIdoublet{} and \CIsinglet{} observed in Figures \ref{fig:CI9850_results} \& \ref{fig:CI8727_results}. On the other hand, taking the conservative departure track of he3p3 could be an underestimation for the SESNe in our sample, which would cause us to underestimate their temperatures.

\section{Discussion}
\label{sec:discussion}

The picture that arises from the previous section is a mixed one. On the one hand, comparisons of \CIdoubletdiag{} and \CIsingletdiag{} indicate that our models significantly overproduce carbon emission (an issue which has also been present in earlier SN modeling efforts \citep{Fransson_1989_Canon, Mazzali_2010_2007grcarbon, Jerkstrand_2015_Canon, Dessart_2021_SESNespectra}). On the other hand, the resulting \Mcarbon{} estimates discussed in Section \ref{subsec:carbon_mass_estimation} seem to be well reproduced by our model grid. We stress once more that these estimates are a lot more uncertain than those of \CIdoubletdiag{} and \CIsingletdiag{}, due to the added uncertainties in NLTE departure and distance and extinction towards the host. 

We will now discuss multiple aspects of stellar evolution that may influence the final carbon mass in the ejecta.

\subsection{Stellar Evolution}
\label{sec:stellar evolution}

A first question to consider is how the carbon yields in the stellar models from \citet{Woosley_2019_inputmodels} adopted here compare to other contemporary pre-supernova grids available in the literature. In Table \ref{tab:stellar_evolution_comp}, we compare our four progenitors to binary-stripped progenitors in \citet{Farmer_2021_Carbonintroduction}. We compare each of our progenitors to the model with the most similar \MpreSN{} in \citet{Farmer_2021_Carbonintroduction}. \footnote{We choose \MpreSN{}, as this has been shown to be the driving parameter for the final nebular spectra that the SN produces \citep{Dessart_2023_grid}}

\begin{table}
    \centering
    \begin{tabular}{c|c|c|c}
    \hline
    model & \MpreSN & \Mzams & \Mcarbon \\ 
    & [\Msun{}] & [\Msun{}] & [\Msun{}] \\ \hline \hline
    
    he3p3    & 2.67  & 16.1  & 0.053 \\
    \citet{Farmer_2021_Carbonintroduction}  & 3.19  &  11 &  0.055  \\ \hline
    he4p0    & 3.15 & 18.1 & 0.09       \\
    \citet{Farmer_2021_Carbonintroduction}  & 3.19  & 11 &  0.055   \\ \hline
    he6p0    &  4.45  & 23.3 & 0.24       \\
    \citet{Farmer_2021_Carbonintroduction}  & 4.42 & 14 &  0.084   \\ \hline
    he8p0    &  5.64  & 27.9 & 0.486      \\
    \citet{Farmer_2021_Carbonintroduction}  & 5.61 & 17 &  0.198   \\
    \end{tabular}
    \caption{Comparison between the ejecta models used in this work, and those simulated in \citet{Farmer_2021_Carbonintroduction}. We compare the models in our grid to those with closest \MpreSN{} in the \citet{Farmer_2021_Carbonintroduction} grid. Again, \Mcarbon{} is here the carbon mass in the entire ejecta.}
    \label{tab:stellar_evolution_comp}
\end{table}

A first glance immediately makes clear that \Mcarbon{} for the \citet{Farmer_2021_Carbonintroduction} models is consistently lower, being between $\sim$ 0.6 to 0.35 times that of the \citet{Woosley_2019_inputmodels} models. This is a similar difference as to what \citet{Farmer_2021_Carbonintroduction} find when they compare the single star progenitors, which they also simulate, to other single star grids (see their figure 8).

What also becomes apparent, is that while both studies simulate binary-stripped stars, the stars being compared here have very different \Mzams{}. This difference may originate in the treatment of the stripping; where \citet{Woosley_2019_inputmodels} simply removes the entire hydrogen-envelope at the onset of central helium burning, \citet{Farmer_2021_Carbonintroduction} models the stripping by Roche lobe overflow (RLOF) self-consistently. Additionally, both works employ different prescriptions for wind mass-loss, such that the stars in \citet{Woosley_2019_inputmodels} experience stronger mass-loss. So, even though \MpreSN{} is the same for the stars compared here, the conditions influencing their nucleosynthesis during their lives have not been similar.

The simplistic comparisons above already indicate that uncertainties in stellar evolution can yield $\sim$ 50\% difference in ejected carbon mass for similar progenitors. We will now shortly discuss a couple of stellar evolution processes that could alter the mass of carbon present at the time of explosion. 
\subsubsection{Binarity}
\label{sec:discussion_binarity}
For more than a decade now, it has been the consensus \citep{Sana_2012_Canon, Smith_2014_MasslossReview, Moe_2017_PsandQs} that the majority of massive stars are not single stars, but that they rather are in binaries. Motivated by this fact, there have been recent studies (e.g. \citet{Woosley_2019_inputmodels, Laplace_2021_DifferentToTheCore, Farmer_2021_Carbonintroduction, Schneider_2021_MESAsuite}) to investigate the differences in stellar evolution between single and binary stars. Besides the clear difference caused by the loss of the hydrogen envelope through RLOF, an important finding that has been done in \citet{Woosley_2019_inputmodels}, \citet{Laplace_2021_DifferentToTheCore} and \citet{Schneider_2021_MESAsuite} is that binary-stripped stars retain a higher carbon mass fraction \XtwelveC{} at central carbon ignition than (effectively) single stars (see figure 15 in \citet{Laplace_2021_DifferentToTheCore}). 

However, this may not necessarily translate into a higher final \Mcarbon{}; in fact, for \MpreSN{} $\lesssim$ 4.5 \Msun{} (which holds for all our progenitors except the he8p0 model), \citet{Farmer_2021_Carbonintroduction} find that single stars end up with slightly higher \Mcarbon{} than their binary-stripped counterparts. This suggests that the discrepancy in relative carbon luminosities between our SESN sample and the stellar models by \citet{Woosley_2019_inputmodels} can not be explained through a large part being single star progenitors.

Finally, one can consider the reduction of carbon in the final ejecta through RLOF of the helium envelope. For helium stars with \Mi{} $\lesssim$ 3.5 \Msun{}, the helium envelope is expected to expand upon core helium depletion (see e.g. figure 4 in \citet{Woosley_2019_inputmodels}). For sufficiently close binaries ($P_{\text{orb}}$ $\lesssim$ 10 -- 20d \citep{Tauris_2015_Ultrastripped, Yoon_2017_Binary_Massloss}), this expansion ushers in a second phase of RLOF, stripping the He/N layer and significant parts of the He/C envelope. Reminding the reader of Figure \ref{fig:hec_vs_oc_flux_ratio}, this could lead to significant carbon luminosity reductions. This expansion did not occur in our model grid. However, the majority of the SNe in our sample are not expected to originate from such low mass progenitors (see e.g. table 2 in \citet{Barmentloo_2024_Nitrogen}), and the sample of objects with necessary ejecta masses of $\lesssim$ 0.5 -- 1.0 \Msun{} found in the literature is also small.

\subsubsection{Rotation}
Another important parameter in stellar evolution is rotation \citep{Langer_1992_Rotation, Heger_2000_Rotation, Meynet_2000_Rotation}, as the mixing it induces increases the extent of the core burning region.  \citet{Limongi_2018_SNmodels} studied rotation\footnote{Note that the stellar evolution code is in 1D, so that rotation is treated using a mixing efficiency parameter. See their section 3 for details on this calibration} for a large grid of single stars. In this study, three different values for the initial rotational velocity ($v =$ 0, 150 and 300 \kms{}) were considered. The authors find that for the lower mass stars (\MCOcore{} $\lesssim $ 5 \Msun{}), increased rotation leads to a significant decrease in central carbon abundance at central carbon ignition (\XtwelveC{} drops from $\sim$ 0.3 for $v = 0$ \kms{} to $\sim$ 0.05 -- 0.2 for $v = 300$ \kms{}; see their figure 19). Interestingly, the final carbon yields \textit{increase} with rotational velocity (see their table 8). 

As the grid in \citet{Limongi_2018_SNmodels} only consists of single stars, and grids of rotating massive stars in binaries are lacking, comparing directly to our adopted stellar models (which are non-rotating) is not possible. However, except for the tightest binaries (which lose their envelopes already on the MS), MS-evolution of single and binary stars should be very similar, and one would thus expect the effect of rotation on the MS to also occur for stars in binaries. Most importantly, including rotation would lead to larger helium cores for the same \Mzams{} in our grid. 

\subsubsection{Metallicity}

\citet{Ma_2025_CarbonFromBinaryModels} recently presented a study where they investigated the effect of metallicity on carbon yields from massive stars (both single and in binaries), by simulating the same stars for SMC, LMC and solar metallicity. They find that the influence of metallicity is dominant, being larger than the effect of e.g. binaries or other stellar parameters, with stars producing relatively more carbon with increasing metallicity. It should be noted that these increased yields only occur from \Mzams{} $\gtrsim$ 18 \Msun{} (or \MCOcore{} $\gtrsim$ 4.5 \Msun{} ), so that a non-negligible fraction of our sample would be excluded (see e.g. the progenitor mass estimates in \citet{Barmentloo_2024_Nitrogen}. 

Recent studies of CCSNe metallicities \citep{Anderson_2010_MetallicityCCSNe, Ganss_2022_MetallicityCCSNe, Xi_2024_MetallicityCCSNe} find the quantitiy 12 + log(O/H) to range between 8.1 and 8.7, with mean metallicities of 8.4 -- 8.5 and 68\% of the population within 0.1 -- 0.15 from the mean. Assuming a solar value of 12 + log(O/H) = 8.69 \citep{Asplund_2021_canon}, this means that the majority of CCSNe have metallicities of $\sim$ 0.4 -- 1.0 $Z_{\odot}$. The solar metallicity models from \citet{Woosley_2019_inputmodels} are thus likely to have a somewhat higher metallicity than typical in our observed sample, which especially for higher (\Mzams{} $\gtrsim$ 18 \Msun{}) mass progenitors could lead to overestimation of carbon yields.

\subsubsection{Shell Mergers}

Another mechanism that might decrease carbon abundance in massive stars is a C-O shell merger \citep{Rauscher_2002_COmerger, Meakin_2006_ShellmergerMultiD, Rizzuti_2024_shellmerger}. In this process, which occurs in the final days of the star's life, carbon and neon are ingested into the oxygen burning shell. Among other things, a C-O shell merger leads to the destruction of the carbon in the inner half of the O/C-layer, and would thus have a similar order effect on carbon emission as molecules. While the mechanism has been found in both 1D and multidimensional simulations, there remains large uncertainty on how common these mergers are, especially because observational evidence is still lacking. A recent study by \citet{Roberti_2025_COmergerssample} found that for a large collection of single-star literature models, $\sim$ 20\% experience a C-O shell merger, with the majority having low \MCOcore{} ($\lesssim$ 5 \Msun{}) and central \XtwelveC{} ($\lesssim$ 0.28) at the end of core helium burning. The stellar models that we studied in this work experienced no such merger, which can largely be explained by the fact discussed before in Subsection \ref{sec:discussion_binarity}, namely that binary-stripped stars have higher \XtwelveC{} than their single star counterparts. In fact, the four models studied in this work all have \XtwelveC{} $>$ 0.28. A similar argument can be made for the C-Ne shell mergers observed in the single-star models by \citet{Laplace_2025_CNeshellmergers}; in their simulations, these occur for models with low \XtwelveC{} (0.17) and masses above the first compactness peak (\MCOcore{} $\gtrsim$ 6.5 \Msun{}), a parameter space which is not covered by the stellar models in our grid. All in all, we deem it thus unlikely shell mergers are relevant for a significant part of our sample.

\subsection{He/C-zone extent from nebular SESNe spectra}
\label{subsec:hec_size}

\begin{figure}
    \centering
    \includegraphics[width=0.98\linewidth, height = 1.36 \linewidth]{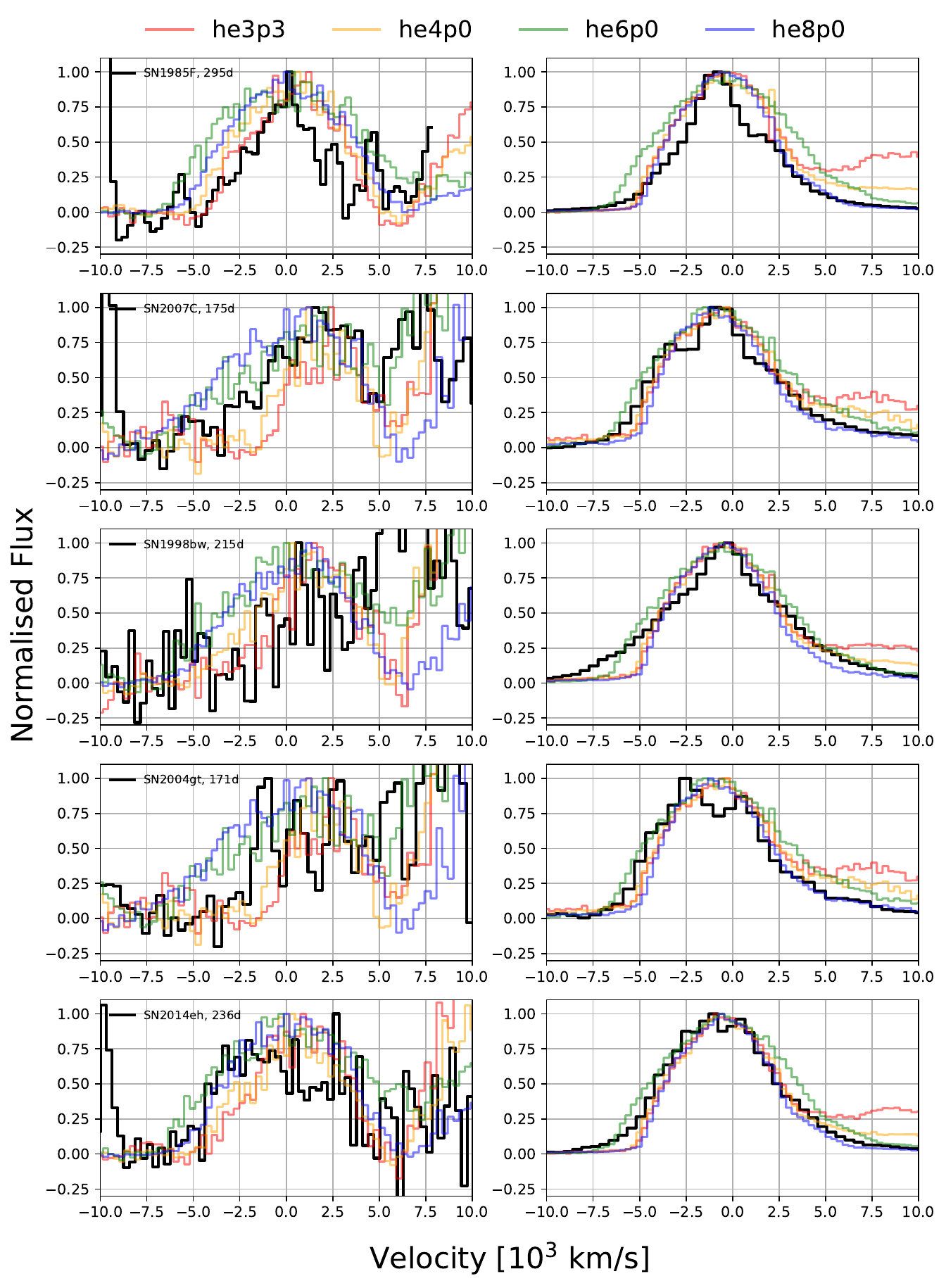}
    \caption{A plot comparing line widths of \CIdoublet{} (left half of the figure) and \OIdoublet{} (right half of the figure) for five out of nine SNe in our sample with observed \CIdoublet{} signal. Observations are indicated in black, models are indicated in colour. For the model spectra, the epoch closest to the epoch of the observed spectrum was selected. The x-axis zero-points were set as 9844 Å and 6316 Å, respectively. The fluxes were normalised to the peak of the respective lines studied. As discussed in Section \ref{subsec:hec_size}, model line widths for \CIdoublet{} are noticeably larger than for \OIdoublet{}, while the figure shows that this does not hold for all observations. This is an indication for a smaller He/C zone in (some) observed SNe than in the models we adopted.}
    \label{fig:hec_sizea}
\end{figure}

\begin{figure}
    \centering
    \includegraphics[width=0.98\linewidth, height = 1.12 \linewidth]{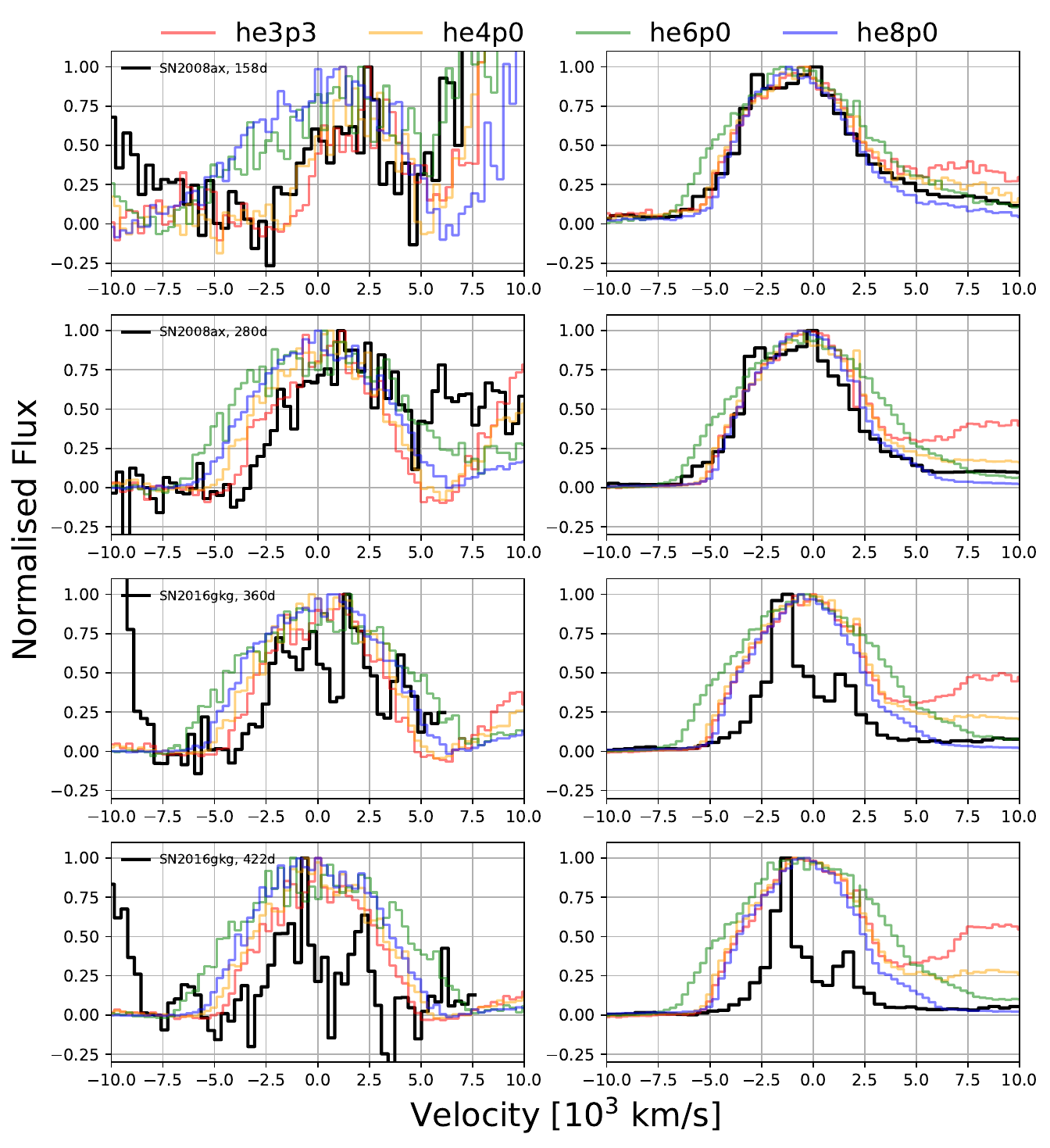}
    \caption{Same as Figure \ref{fig:hec_sizea}, but for the remaining four \CIdoublet{} detections.}
    \label{fig:hec_sizeb}
\end{figure}

As was noted in Section \ref{sec:results_8727}, Figure \ref{fig:CI8727_results} shows that the Type Ic SNe  have on average the highest \CIsingletdiag{} values in our sample. However, in the canonical picture Type Ic SNe lack a helium layer, and thus a He/C layer. Naively, we would thus expect them to have the lowest relative carbon luminosities. 

One potential explanation for this discrepancy could be that the He/C extent in the \citet{Woosley_2019_inputmodels} and \citet{Ertl_2020_explosions} models is larger (or has a larger carbon abundance) than typical in the universe. Between different stellar evolution models, the characteristics of this zone vary appreciably, as its formation is influenced by many uncertain factors, such as rotation and (wind- and binary) mass loss (see sections 4.2 and 5.1 in \citet{Limongi_2018_SNmodels}). For example, while the ratio of the O/C to He/C zone masses for our models is similar to those studied in \citet{Jerkstrand_2015_Canon}, the carbon abundance in the He/C zone of their 17 \Msun{} model is 0.045, where it is 0.39 in our he8p0 model (which comes closest in ejecta mass). 

While determining the abundance of carbon in the He/C-zone from the carbon lines themselves is not possible, their line widths might turn out to be of use. As was noted by \citet{Fransson_1989_Canon}, carbon lines in nebular spectral synthesis models turn out wider than other strong nebular lines such as e.g. \OIdoublet{} and \CaIIdoublet{}. The reason for this is that the carbon lines originate partly from the He/C-layer, which has a higher velocity than the macroscopically mixed core. The more the carbon line width thus differs from e.g. the \OIdoublet{} line width, the larger the extent of the He/C-layer. While the complex blend of \CIsinglet{} makes it complicated to determine line widths accurately, the effectively isolated \CIdoublet{} serves as a great tool for such a comparison.

In Figures \ref{fig:hec_sizea} and \ref{fig:hec_sizeb}, we compare the line widths of \CIdoublet{} and \OIdoublet{} with the model line widths, for our nine spectra with a \CIdoublet{} detection. Comparing line widths is complicated for \OIdoublet{} in cases of blending on the red side with \NIIdoublet{} (mostly late epoch, low mass models), and for \CIdoublet{} in cases of blending on the red side with \ion{Co}{II} (mostly early epoch, low mass models). However, some insights can still be gained. For example, in both observations of SN 2008ax the \OIdoublet{}-profile compares well to the he3p3, he4p0 and he8p0 models, which all have $V_{\text{core}} \sim$ 4 500 \kms{}. At the same time, the \CIdoublet{}-profile is noticeably more narrow than any of the models. This suggests that in SN 2008ax, the extent of the carbon emitting material is smaller than in our model stars. The same thing can be said, at least for the he8p0 model (which has the comparatively largest He/C-layer) for all other spectra (except for SN 2016gkg), and in many cases for the he6p0 model as well. For models other than he8p0, line width differences are not as pronounced and thus harder to observe, which is further complicated by the poor S/N of many of the \CIdoublet{} observations. This fact combined with the small sample size makes drawing any general conclusions difficult, but these comparisons do indicate that the He/C-layer extent for some SNe may indeed be smaller than in the stellar models studied in this work. Furthermore, they show once again that high quality observations of \CIdoublet{} during the nebular phase should be a focus of future SN observing campaigns.

\section{Conclusions}

In this work, we presented the first systematic study of carbon emission lines in SESNe in the nebular phase. We created a grid of \SUMOcodename{} models based on \texttt{KEPLER} helium star models, and found \CIsinglet{} and \CIdoublet{} to be the only carbon lines of significance. For this grid, we studied the effects on carbon emission of uncertain input physics, namely the filling factors of the core zones and the effect of molecule formation. Next, we compared the carbon line luminosities in a sample of SESNe nebular spectra to our model grid. To extract \LCIsinglet{}, a line that is part of a complex blend of at least six emission lines, we created the \codename{} code, a trimmed-down version of \SUMOcodename{} that focuses solely on the line formation physics for this blend. Our main findings are as follows:

\begin{enumerate}[I]

    \item In our model grid, carbon lines are only effectively formed in the O/C and He/C zones. The relative importance of these zones varies with helium core mass, with increasing core mass leading to increased importance of the He/C zone. This is caused by an increased ratio of He/C to O/C zone mass, as well as a heightened neutral carbon fraction in the He/C zone.

    \vspace{0.2cm}
    
    \item Carbon line luminosities are only mildly ($\lesssim$ 10$\%$) sensitive to uncertainties in filling factor and molecule formation. We do note that the models in our grid form only a small mass ($\lesssim$ 5 $\times$ 10$^{-6}$ \Msun{}) of carbon monoxide compared to existing SESN literature models ($\sim$ 10$^{-4}$ \Msun{} \citep{Liljegren_2023_MoleculesinSESNe}) and observational estimates ($\sim$ 10$^{-4}$ -- 10$^{-3}$ \Msun{} \citep{Banerjee_2018_COovertoneI, Rho_2021_COmass}). We attribute this low mass to a comparatively small mass of the O/C zone in our ejecta models, as well as lower densities in this zone. 

    \vspace{0.2cm}
    
    \item The relative strength of \CIdoublet{} compared to the optical spectrum, \CIdoubletdiag{}, is found to increase strongly with time, from $\sim$ 1 at 150d up to 10 -- 20 at 400d. Furthermore, \CIdoubletdiag{} is found to increase with helium core mass, albeit not monotonically. Combined with the fact that in all models and at all epochs \CIdoublet{} is well-treated as an LTE line, this makes \CIdoublet{} an extremely useful line to study carbon nucleosynthesis in SNe, especially when observed at epoch $\gtrsim$ 300d. 
    
    For the majority of spectra in our sample, the wavelength region of \CIdoublet{} was either not included in the spectrum, or the quality of the spectrum was too poor. Nonetheless, when comparing the relative luminosity \CIdoubletdiag{} in models to our observational sample, we find that more than half of observed spectra have \CIdoubletdiag{} below even our lowest model, indicating an overestimation of \CIdoublet{} emission in our models. 

    \vspace{0.2cm}
    
    \item The evolution of \CIsingletdiag{} in the model grid is not as pronounced in time or mass, decreasing from $\sim$ 15 to $\sim$ 10 between 150d and 400d. Importantly, we also find that \CIsinglet{} is not well treated as an LTE line, with departure coefficients at 150d already down to $\sim$ 0.7 -- 0.9, decreasing to 0.15 -- 0.2 at 400d. Comparing the model grid values with the estimates of \CIsingletdiag{} obtained by our newly developed \codename{} code, we again find most observations to have lower values than the model grid. The discrepancy is the largest for Type IIb SNe and still clear for Type Ib SNe. For Type Ic SNe, observations are not incompatible with the model values. 

    \vspace{0.2cm}

    \item We provide estimates and upper limits of the neutral carbon mass through an analytical, LTE formalism. However, the non-LTE nature of \CIsinglet{} leads to our LTE formulae largely overestimating these masses. The added uncertainties in host extinction and distance to the SN make these mass estimates highly complicated. Therefore, we advise readers to only use this formalism at early epochs (where the LTE departure is smallest), and for objects with well known host extinction and distance. In fact, the higher densities in their ejecta and consequently smaller LTE departures should make Type II SNe significantly better candidates to apply the formalism to than SESNe, and we therefore encourage such a study.

    \vspace{0.2cm}

    \item Multiple potential explanations for the found discrepancies in \CIsingletdiag{} and \CIdoubletdiag{} are discussed, including stellar evolution modeling, binary fractions, rotation, metallicity and C-O shell mergers. From our limited analysis, we find that there are still appreciable differences in carbon yields for different stellar evolution codes, and at the same time we deem metallicity and shell mergers to be unlikely causes of the discrepancy. 

    \vspace{0.2cm}
    
    \item When studying the line widths of \CIdoublet{} in observed spectra, we find indications that they are not significantly different from the line widths of \OIdoublet{}. This hints that the extent of the He/C zone may be smaller than expected from models, where line widths of \CIdoublet{} are clearly larger than those of \OIdoublet{}. However, the limited sample of detected \CIdoublet{} in observations does not allow us to conclude anything definitively.

\end{enumerate}

Based on the conclusions presented here, we would like to close off this work by joining a call to observers that was also present in the recent work by \citet{Liu_2025_SN2022pul}, i.e. to put increased attention into obtaining high quality \CIdoublet{} observations of SNe. Its LTE nature makes it a comparatively simple line to model and analyse, and the close to monotonic \updated{increase of \CIdoubletdiag{}} with helium core mass for a given epoch also gives it potential as a progenitor mass diagnostic. Furthermore, its linewidth can be used as a probe for the extent of the He/C zone, which in turn can provide valuable information to stellar evolution modelers. 

\label{sec:conclusion}

\section*{Acknowledgements}

This research was funded by the Swedish Research Council (Starting Grant 2018-03799, PI: Jerkstrand). The computations in this work were enabled by resources provided by the Swedish National Infrastructure for Computing (SNIC), the National
Academic Infrastructure for Supercomputing in Sweden (NAISS),
and at the Parallelldatorcentrum (PDC) Center for High Performance
Computing, Royal Institute of Technology (KTH), partially funded
by the Swedish Research Council through grant agreements no. 2022-06725 and no. 2018-05973.

We thank Stefano Valenti for kindly providing the full spectra (out to $\gtrsim$ 10 000 Å) for SN 2009jf and SN 2007gr. For a crash course in the usage and interpretation of the CFD3 database, we thank Alex Pedrini. For insights in host extinction measurements, we thank Avinash Singh. We thank Bart van Baal, Anamaria Gkini and Yang Hu for useful discussions.

This research has made use of the NASA/IPAC Extragalactic Database, which is funded by the National Aeronautics and Space Administration and operated by the California Institute of Technology.

\section*{Data Availability}

In this work, nebular phase SESNe spectra were used that were easily obtained with the help of the web-tools WISeREP (\url{https://www.wiserep.org}) and the OSC (\url{https://github.com/astrocatalogs/supernovae}. The stellar evolution and explosion models from \citet{Woosley_2019_inputmodels} and \citet{Ertl_2020_explosions} were obtained from \url{https://wwwmpa.mpa-garching.mpg.de/ccsnarchive/}. The code necessary to produce the figures in this manuscript is available at \url{https://github.com/StanBarmentloo/carbon_emission_in_SESNe}. The \codename{} package is made publicly available and can be found at \url{https://github.com/StanBarmentloo/CaNARY}.)

\bibliographystyle{mnras}
\bibliography{references}

\appendix

\section{Determining Upper Limits on \LCIdoublet{}}
\label{app:doublet_luminosity}

As discussed in Section \ref{subsec:estimating_9850}, the majority of observed spectra showed no significant \CIdoublet{} emission, while some did have a clear signal. Here we outline our approach in both of these cases.

\subsection{Case 1: \CIdoublet{} is observed}
\label{app:doublet_is_observed}

\begin{figure}
    \centering
    \includegraphics[width=0.475\textwidth]{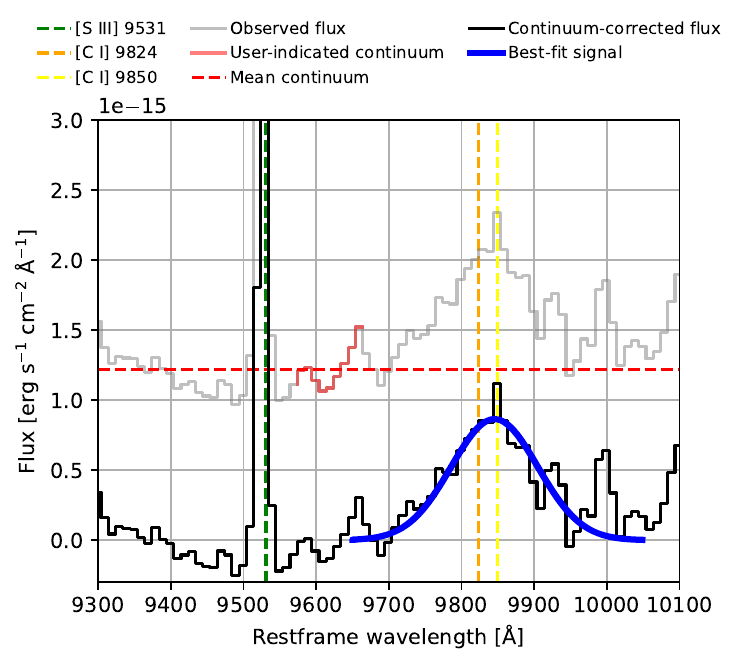}
    \vspace{-0.4cm}
    \caption{An example of fitting \LCIdoublet{} for SN 1985F. The selected continuum region covers 50 Å on both sides of 9620 Å. The mean flux in this region is removed, resulting in the spectrum in black. To this spectrum, a simple Gaussian fit is performed. In this example, the best-fit parameters for the Gaussian are $A = 8.65 \times 10^{-16}$ \fluxunit{}, $\mu = 9846$ Å, $\sigma $ = 59.6 Å .}
    \label{fig:estimating_9850_example}
\end{figure}

When \CIdoublet{} is observed, \codename{} prompts the user to indicate a local minimum in the flux levels and its width. The mean flux in the selected region is treated as a flat continuum and subtracted from the spectrum (see e.g. Figure \ref{fig:estimating_9850_example}). Next, the spectrum is rebinned to a resolution of 10 Å to increase S/N. Finally, a single Gaussian is fit to the remaining line profile and integrated to obtain \LCIdoublet{}. To estimate the uncertainty on this integrated flux, we first determine the standard deviation $\sigma$ of the flux regions adjacent to (but not including) the signal region. Then, the uncertainty estimate is defined as the ratio of $\sigma$ and the amplitude of the best fit Gaussian, multiplied by the total integrated flux under the Gaussian.

\begin{figure}
    \centering
    \includegraphics[width=0.475\textwidth]{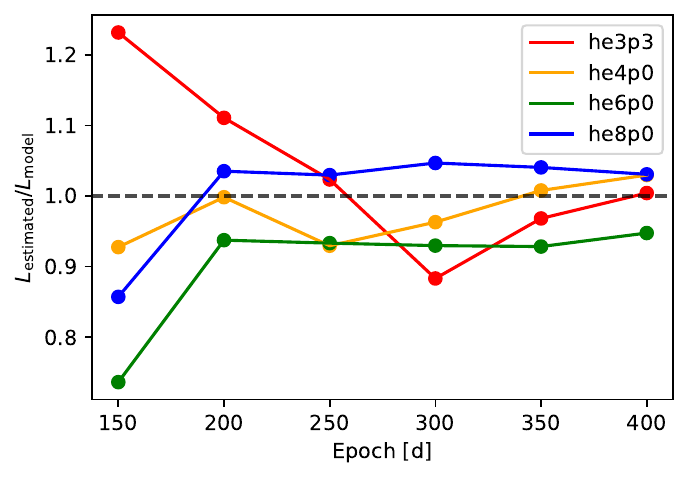}
    \caption{Comparison between \LCIdoublet{} estimated through our methodology (described in Appendix \ref{app:doublet_is_observed}) and the \LCIdoublet{} in our four model spectra. }
    \label{fig:validating_9850}
\end{figure}

To validate how well our fits capture the actual \LCIdoublet{}, we can use the described methodology on our \SUMOcodename{} models, where we know the exact value of \LCIdoublet{}. This comparison is of great importance to check how well we can capture \LCIdoublet{} at early times and for low masses, when contamination of \ion{Co}{II} on the red side could mislead the fitting. The comparison between 'true' and estimated \LCIdoublet{} are shown in Figure \ref{fig:validating_9850}. Typical errors are on the 5\% level, except for the earliest epoch, where they increase to 20\%. Reassuringly there is no preference for over or underestimation. 

\subsection{Case 2: \CIdoublet{} is not observed}
\label{app:doublet_is_not_observed}

In the case when \CIdoublet{} is considered not observed, we follow the prescription used in \citet{Shappee_2013_UpperLimit} and \citet{Lundqvist_2015_UpperLimit}. The user is first requested to define a region that includes both the expected signal region, as well as some surrounding background. This region is then smoothed with a second-order Savitzky-Golay polynomial with window-size of 300 Å. The resulting smoothed function is removed as a continuum from the original unsmoothed spectrum, leaving a so-called 'net spectrum'. This net spectrum was rebinned into 10 Å bins to increase S/N. 

With the net spectrum obtained, the determination of the upper limit can properly commence. First, we determine the 1$\sigma$ noise level in the background, with the background defined as the regions where $\lambda < 9704$ Å or $\lambda > 9984$. These values were chosen symmetrically around the theoretical line centre as limits on where one would expect any signal from unseen \CIdoublet{} to potentially be present. For the flux in the two background regions the standard deviation is then determined and averaged (lets call this $\sigma_{\text{bg}}$). Now we create $n = $ 10.000 iterations of a Gaussian with amplitude $A$, centroid $\mu = 9844 Å$ and width $\sigma = 70$ Å, and for each iteration add a new instance of white noise with a standard deviation $\sigma_{\text{bg}}$. For each iteration $n_{i}$ we determine the total integrated flux in the spectral region, $L_{i}$. By picking $L_{1587}$ and $L_{8413}$ in the ordered $L_{i}$ list and taking the difference between them, one then obtains an effective 1$\sigma_{L}$ uncertainty on \LCIdoublet{}. The algorithm continues to increase the amplitude $A$ of the injected Gaussian, until $L_{5000}$ = 3$\sigma_{L}$ (with $L_{5000}$ the median of the ordered list of $L$). The integrated flux of a Gaussian signal with the amplitude $A_{3\sigma}$ for which this holds is then interpreted as the 3$\sigma$ upper limit on \LCIdoublet{}. 

\begin{figure*}
    \centering
    \includegraphics[width=0.95\textwidth]{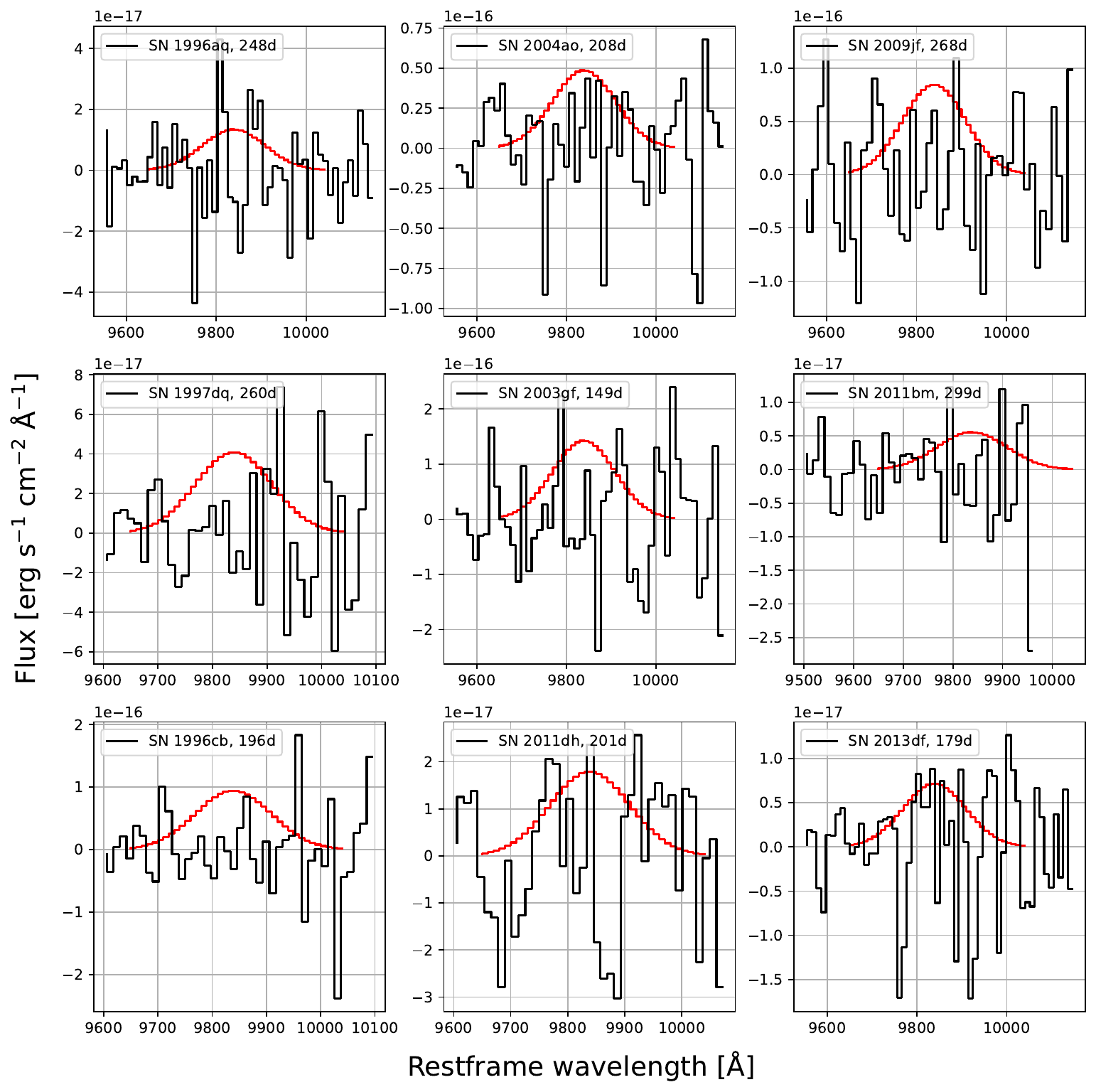}
    \caption{A representative selection of the upper limits determined in Appendix \ref{app:doublet_is_not_observed}. Each spectrum has had a continuum removed, that was determined using a Savitzky-Golay filter (see the corresponding section). The 3$\sigma$ upper limit signals are shown in red.}
    \label{fig:validating_9850_upperlimits}
\end{figure*}

Validating these upper limits is a non trivial undertaking, as the whole premise is that there is no signal to be checked. What we can do is judge by eye whether or not the limits look credible. A collection of upper limit signals we obtained is presented in Figure \ref{fig:validating_9850_upperlimits}. What becomes clear is that for some objects the upper limit seems to be too conservative, where for others it might be somewhat too generous. An important factor in the object to object variations is the wavelength range of the spectographs used; for many of our spectra the spectrum ends somewhere between 10.000 to 10.200 Å, putting the right hand background region at the very end of the range. This typically leads to much worse S/N in the redward background region than in the blueward background region. Depending on the exact range, this can cause an over- or underestimation of $\sigma_{bg}$ and in turn of $A_{3\sigma}$. All in all, we believe that the examples shown in Figure \ref{fig:validating_9850_upperlimits} are acceptable upper limits, but at the same time we stress that they can sometimes be too generous.

\section{Simulating \CaNIRtriplet{} line profiles}
\label{app:singlet_luminosity}

Here we describe the \codename{} code described in Section \ref{subsec:estimating_8727} in further detail. The code tracks a total of $10^{6}$ photons from creation to escape. 

\subsection{Photon Creation}
\label{app:photon_creation}

When a photon is created, it has an initial wavelength $\lambda_{\text{init}}$, radius $r_{\text{init}}$ and angle $\mu_{\text{init}}$. 

The wavelength $\lambda_{\text{init}}$ is drawn randomly from the six lines in Table \ref{table:six_lines}. \footnote{In our simulations, the line Fe I $\lambda$ 8688 also showed non negligible optical depths, but we do not include it in the \codename{} modeling. This choice was made as we wanted to keep \codename{} as simplistic as possible, and the potential minor gain in accuracy did in our eyes not justify the increase in complexity. } The probability of each line's wavelength being selected is a free parameter and should be understood as the line's luminosity generated by thermal processes. 

\begin{figure}
    \centering
    \includegraphics[width=0.475\textwidth]{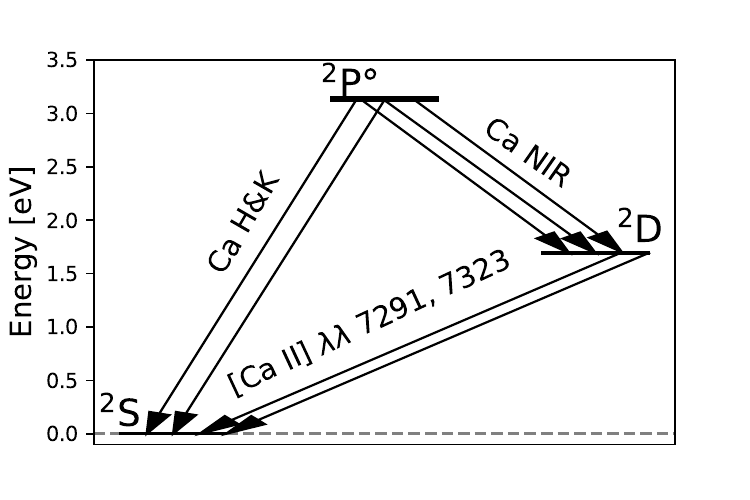}
    \caption{A diagram showcasing the three lowest \ajc{terms} in the \ion{Ca}{II} ion, and the optical + NIR transitions connecting them.}
    \label{fig:calcium_level_structure}
\end{figure}

We remove two of the six free parameters here by always assuming the same ratio of luminosities for the three individual \CaNIRtriplet lines. We justify this as follows. The luminosity of a line, \ajc{per unit volume}, in the Sobolev approximation \citep{Sobolev_1957_Canon} is given by Equation \ref{eq:luminosity_def}:
\begin{equation}
    L = n_{u} A \beta_{S} h\nu,
\label{eq:luminosity_def}
\end{equation}
where $n_{u}$ is the number density in cm$^{-3}$ of the upper level, $A$ is the transition rate in s$^{-1}$ and $\beta_{S}$ is the escape probability in the Sobolev approximation. For an optically thick line (which the \CaNIRtriplet{} lines are at all times, see Section \ref{subsec:estimating_8727}), $\beta_{S} \sim 1/\tau_{S}$ \citep{Jerkstrand_2017_Book}, which in turn is given by (\ajc{ignoring the stimulated emission correction}):
\begin{equation}
    \tau_{S} = \frac{1}{8\pi} \frac{g_{u}}{g_{l}}A\lambda^{3}n_{l} t.
\label{eq:tau_def}
\end{equation}
Inserting Equation \ref{eq:tau_def} into \ref{eq:luminosity_def} and rearranging, we find that:
\begin{equation}
    L = \frac{8\pi hc}{\lambda^4t} \frac{n_{u}}{n_{l}} \frac{g_{l}}{g_{u}}. 
\end{equation}
In other words, the ratio of luminosities for individual lines in the triplet is given by (ignoring the factor $\left(\frac{\lambda_{\text{line1}}}{\lambda_{\text{line 2}}}\right)^{-4}$, which is close to unity as the wavelengths are close):
\begin{equation}
\begin{aligned}
    \frac{L_{8498}}{L_{8542}} = \frac{n_{3}}{n_{2}} \frac{g_{2}}{g_{3}}, \\
    \frac{L_{8498}}{L_{8662}} = \frac{n_{5}}{n_{4}} \frac{g_{4}}{g_{5}}, \\
    \frac{L_{8542}}{L_{8662}} = \frac{ \frac{n_{5}}{n_{4}} \frac{g_{4}}{g_{5}} } {\frac{n_{3}}{n_{2}} \frac{g_{2}}{g_{3}}}.
\end{aligned}
\label{eq:luminosity_ratios}
\end{equation}
Here it is important to realise from what levels the \CaNIRtriplet{} lines originate (see Figure \ref{fig:calcium_level_structure}). The Ca II ion has a singlet ground state (referred to as $^{2}\text{S}$ in Figure \ref{fig:calcium_level_structure}), followed by a doublet ($^{2}\text{D}$) and another doublet ($^{2}\text{P}°$). If one assumes that statistical equilibrium for levels within a multiplet is obtained faster than the multiplet is depopulated, one can say that $n_{5}/n_{4} = g_{5}/g_{4} \times e^{-\frac{E_{5} - E_{4}}{k_{B}T}}$ and $n_{3}/n_{2} = g_{3}/g_{2} \times \frac{E_{3} - E_{2}}{k_{B}T}$. Plugging this result into Equation \ref{eq:luminosity_ratios}, one finds that all components of the triplet have the same luminosity (barring the Boltzmann factor, which is close to unity).

\ajc{For levels 2 and 3, this equilibrium holds to low densities, because the radiative deexcitation can occur by forbidden lines only. Thus, it is robust to take $L_{8498}/L_{8542}=1$. Levels 4 and 5 are,  however, radiatively deexcited by allowed transitions. While the transitions are generally optically thick, the resulting $A\beta_S$ values are still larger than the collisional depopulation rates, so equilibrium cannot be assumed. Some subtlety now enters the argument. The emissivity ratio we are seeking refers to emission following populating processes other than absorption of triplet lines themselves (that process is what the code simulates). The two main drivers for this are thermal excitation (from ground and/or $^2$D) and HK-line absorption. The first process tends to dominate at early nebular phases, the second later on. Starting with the second, because the HK lines have the same A-value, the photoexcitation flows through them to $n=4$ and $n=5$ will tend to be similar 
, as long as the diffuse UV field does not strongly differ blueward of 3933 Å compared to 3968 Å. For thermal excitation, the twice as high statistical weight of level 5 means it will receive twice as high flow as level 4. Thus, for the HK fluorescence case a ratio of 1 for $L_{8498}/L_{8662}$ is the best choice (same as in equilibrium), and for thermal excitation case a ratio of 2 is the best choice. Tests were carried out showing that choosing 2 instead of 1 did not significantly impact the C I luminosity estimate.}

\vspace{0.2 cm}

\begin{table}
\begin{tabular}{l|l}
Line           & Free parameters  \\
\hline
\hline
O I 8446       & $L_{\text{\ion{O}{I}}}$, $\tau_{\text{\ion{O}{I}}}$        \\
\hline
Ca II 8498     & $L_{\text{\ion{Ca}{II}}}$           \\
\hline
Ca II 8542     & $L_{\text{\ion{Ca}{II}}}$            \\
\hline
Ca II 8662     & $L_{\text{\ion{Ca}{II}}}$            \\
\hline
{[}C I{]} 8727 & $L_{\text{[\ion{C}{I}]}}$             \\
\hline
Mg I 8806      & $L_{\text{\ion{Mg}{I}}}$, $\tau_{\text{\ion{Mg}{I}}}$        \\
\hline
               & $V_{\text{out}}$ \\
\end{tabular}
\caption{A summary of the six lines which have been identified to make up the \CaNIRtriplet{}-complex. The second column indicates the free parameters associated with each line in our \codename{} code (see Section \ref{subsec:estimating_8727} and Appendix \ref{app:singlet_luminosity}).}
\label{table:six_lines}
\end{table}

The radius $r_{\text{init}}$ is drawn from a sphere with uniform density, resulting in a parabolic line profile. Such a distribution is supported by observations, which suggest that the core of the SN ejecta (where all our simulated elements are located) is macroscopically mixed \citep{Fransson_1989_Canon, Shigeyama_1990, Ennis_2006_CassiopeiaAmixed}. Observed nebular SN line profiles are usually a mix of a parabolic and Gaussian line shape. We tested both line profiles, but found that for the majority of the SNe in our sample, Gaussian line profiles resulted in significantly worse matches. 

The angle $\mu_{\text{init}}$ is drawn isotropically.

\subsection{Photon Travel}

Once a photon is created, it starts traveling through the nebula at its initial direction, continuously redshifting (in the comoving frame) along the way. If the photon redshifts to one of the six wavelengths in Table \ref{table:six_lines}, it interacts with this line following the Sobolev approximation. Effectively this means that the interaction surface with a line is infinitely thin, so that once a photon redshifts past the line's exact wavelength, it will never interact with it again. To obtain the result of an interaction, all we need to know is the Sobolev optical depth $\tau_{S}$ of the line; at each interaction, there is a probability of $\exp{-\tau_{S}}$ that the photon moves through the line unharmed. In the remaining $1 - \exp{-\tau_{S}}$ cases, the photon interacts with the line. We treat this interaction as an isotropic scattering event. In other words, we assume that the radiative re-emission of the photon occurs on a much shorter timescale than the collisional (de-)excitation to other energy levels within the atom; \ajc{this is based on that $A\beta_S$ for the (allowed) triplet lines are much larger than the collisional deexcitation rates, even when the lines are optically thick}.

In our simulation, we assume the optical depth for the \CaNIRtriplet{} lines to always be $\tau_{S}$ $\gg$ 1. This is supported by our \SUMOcodename{} simulations, where at epochs $<$ 300d\footnote{As already discussed in Section \ref{app:photon_creation}, \codename{} estimates for the regime $\gtrsim$ 300d are less reliable}, $\tau_{S}$ $\gtrsim$ a few in the core. On the other hand, we assume \CIsinglet{} to always be optically thin with $\tau_{S}$ $\ll$ 1. With a transition rate a factor $\sim 10^{7}$ lower than the \CaNIRtriplet{} this is justified, and once again the \SUMOcodename{} simulations show that this assumption holds well for all masses and epochs considered here. For the two remaining lines, we leave $\tau_{S}$ as a free parameter, as they cross the $\tau_{S} \sim 1$ regime within our considered time range. 

Finally, we are left with six free parameters for creating a single simulated line profile. These are given in Table \ref{table:six_lines}. Note how the table lists seven parameters, but that because our code only generates 'relative' luminosities, one effectively eliminates one of the luminosity parameters (e.g. one can put one of the luminosities to 1, and define the others as a ratio to this luminosity).

\subsection{Post Processing \& Model Selection}
\label{app:model_selection}

With the wavelength at escape counted for each photon, a synthetic output spectrum is created. By varying the six free parameters in Table \ref{table:six_lines}, a grid of $\sim$ 3000 such synthetic spectra was simulated. This grid covers $V_{\text{out}}$ values between 3 000 and 7 000 \kms{} at intervals of 500 \kms{}, and relative carbon luminosities \LCIsinglet{} / $L_{\text{tot}}$ from 0 to 0.8, roughly at intervals of 0.05. 

\begin{figure}
    \centering
    \includegraphics[width=0.475\textwidth]{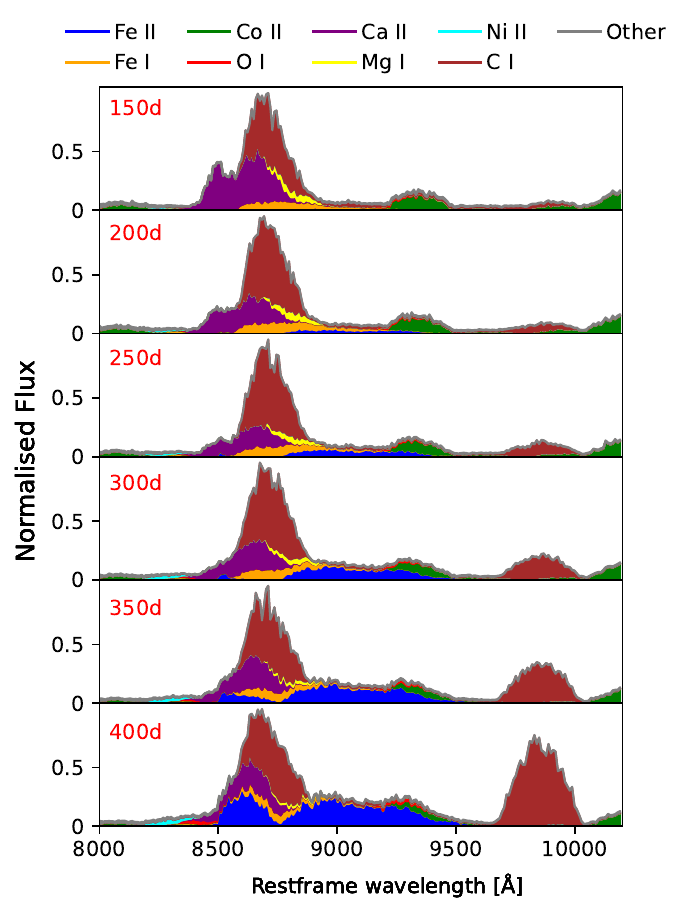}
    \caption{Spectral evolution for the \CIsinglet{} and \CIdoublet{} emission region in the he4p0 model. The most important ions by emergent flux have been indicated in different colours. Where \CIsinglet{} is always present and part of a blend, \CIdoublet{} shows up later but is effectively isolated. }
    \label{fig:ion_by_ion}
\end{figure}

Once the spectral library is created, it is time to compare to the observed spectrum. First, any underlying quasi-continuum in the observed spectrum is removed by simply subtracting the average of the flux levels between 8200 -- 8300 Å and 9000 -- 9100 Å, which during the nebular phase are free of emission lines (see Figures \ref{fig:ion_by_ion} and \ref{fig:spectra_150d} -- \ref{fig:spectra_400d}). The remaining integrated absolute luminosity within the \CaNIRtriplet{}, $L_{\text{CaNIR}}$ is stored, as this will later be used to determine \LCIsinglet{}. Then, the spectrum is normalised to its peak flux to allow comparison to the synthetic spectra. To use as much of the spectral information as possible, the user is then allowed to indicate the following:

\begin{enumerate}
    \item Is there any sign of \MgIeighteightosix{}?
    \item Is there any sign of \OIeightfourfoursix{}?
    \item What is a plausible velocity range for $V_{\text{out}}$?
    \item Within which wavelength range is the \CaNIRtriplet{} contained?
\end{enumerate}

The first three serve to reduce the amount of the spectra that is compared to, preventing as much unnecessary degeneracy during the later fitting; if for example a synthetic spectrum with $V_{\text{out}}$ = 7000 \kms{} happens to give a reasonable fit, but all other spectral lines indicate $V_{\text{out}} \sim$ 4500 \kms{}, question three prevents this model from appearing as good explanation of the actual ejecta structure. The fourth question prevents the scoring from considering regions which are not covered by our modeling; scoring such unmodeled regions anyway results in too high $\chi^{2}$-scores, artificially decreasing the contrast between two model scores.

\begin{figure}
    \centering
    \includegraphics[width=0.475\textwidth]{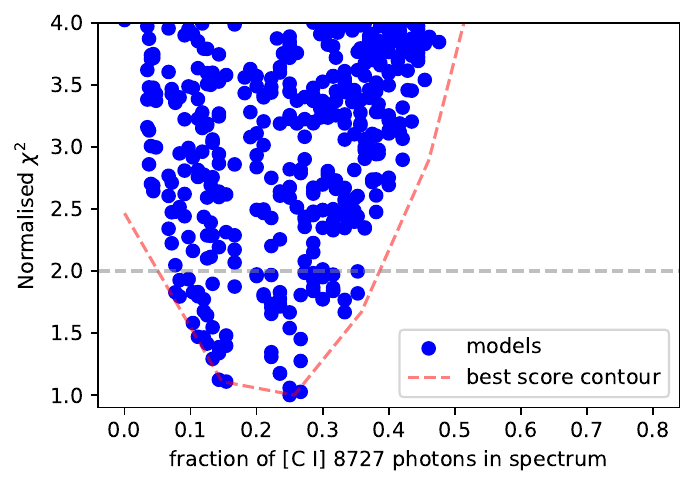}
    \caption{Representative example distribution of normalised $\chi^{2}$ scores (see Section \ref{app:model_selection}) when fitting the \codename{} profiles to observed spectra (SN 2003bg 176d in this case). The distribution is a parabola, allowing for a well defined estimate of the fraction of \CaNIRtriplet{} complex luminosity originating from \CIsinglet{}  }
    \label{fig:chi2_example}
\end{figure}

Now, the remaining subset of synthetic spectra is compared to the processed observed spectrum within the indicated wavelength range by a simple $\chi^{2}$-fit. Each model's score is plotted against its percentage of original \CIsinglet{} photons (for an example, see Figure \ref{fig:chi2_example}). The percentage of the best-fit model is then multiplied with $L_{\text{CaNIR}}$ to finally obtain \LCIsinglet{}. We interpret a $\chi^{2}$-score of 2 as a 1$\sigma$ uncertainty on \LCIsinglet{}. 

\subsection{Validation of the Methodology}

\begin{figure}
    \centering
    \includegraphics[width=0.475\textwidth]{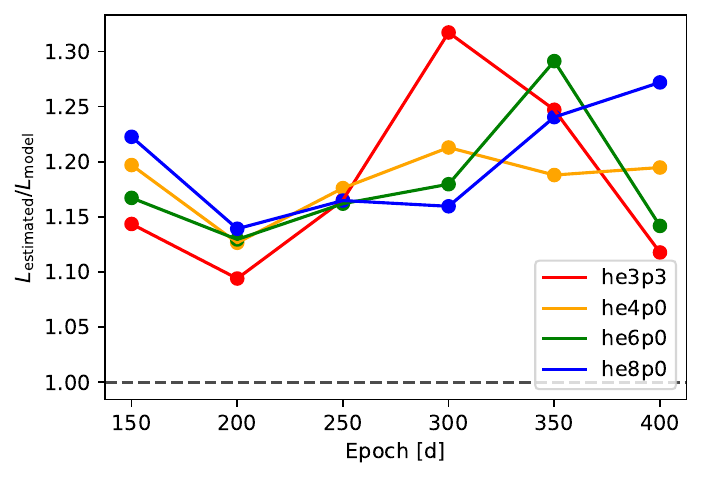}
    \caption{Comparison between \LCIsinglet{}{} estimated through our methodology (described in Appendix \ref{app:singlet_luminosity}) and the \LCIsinglet{} in our four model spectra. }
    \label{fig:validating_8727}
\end{figure}

Just as for \LCIdoublet{}, we applied our methodology to the \SUMOcodename{} models, for which \LCIsinglet{} is known from the simulation. The ratios of \LCIsinglet{} estimated from our methodology to the 'true' \LCIsinglet{} in the models is shown in Figure \ref{fig:validating_8727}. From the comparison we find that \codename{} typically overestimates \LCIsinglet{} by 15 -- 20 $\%$. In the determination of \McarbonI, this effectively leads to an underestimation of the same amount.

An important note to make here is that our models do not cover the entire range of observed \CaNIRtriplet{} line profiles; profiles where the blue peak of the \CaNIRtriplet{} is similar or even greater in strength than the red peak (typical for early epochs of Type IIb SNe) are not found in any of our simulations. This fact is largely expected if our claim of too much \LCIsinglet{} emission in models is taken to be true, but it does mean that we can not check how \codename{} performs for such SNe. We can note however that \codename{} is most sensitive for such objects, resulting in the narrowest confidence intervals for these line profiles.

\section{SESN sample}

In this subsection, we present parameters and references of interest for our two observational samples, namely the 'optical' (Table \ref{tab:optical_sample}) and 'NIR' (\ref{tab:NIR_sample}) samples. 

\begin{table*}
\centering
\begin{tabular}{lllll}
\hline
SN & Type & No. of & Epoch range & Ref \\
 & & Spectra & [d] & \\
 \hline \hline  \\[-.25cm]
1985F & Ib & 1 & 295 & \citet{1985F_a, 1985F_b} \\
1987M & Ic & 1 & 168 & \citet{1987M} \\
1993J & IIb & 20 & 166 -- 326 & \citet{1993J_a, 1993J_b} \\
1996aq & Ib & 2 & 248 -- 290 & \citet{1996aq_1998bw_b} \\
1996cb & IIb & 1 & 196 & \citet{1996cb} \\
1997dq & Ic & 2 & 241 -- 260 & \citet{1997dq_a, 1997dq_b} \\
1998bw & Ic & 3 & 215 -- 391 & \citet{1998bw_a, 1996aq_1998bw_b} \\
2001ig & IIb & 2 & 309 -- 340 & \citet{2001ig_a, 2001ig_b} \\
2003bg & IIb & 5 & 176 -- 301 & \citet{2003bg} \\
2003gf & Ic & 2 & 149 -- 181 & \citet{2003gf} \\
2004ao & Ib & 5 & 172 -- 298 & \citet{2004ao_a, 2004ao_b_2004gq_b} \\
2004aw & Ic & 1 & 249 & \citet{2004aw} \\
2004dk & Ib & 1 & 284 & \citet{2004dk_a_2004gq_a, 2004dk_b} \\
2004gt & Ic & 1 & 171 & \citet{2006ld_b_2007C_b_2007I_b_2008aq_b, Stritzinger_2018_hostreddening} \\
2006ld & Ib & 1 & 141 & \citet{2006ld_a_2007I_a, 2006ld_b_2007C_b_2007I_b_2008aq_b} \\
2007C & Ib & 2 & 141 -- 175 & \citet{2007C_a_2008bo_a, 2006ld_b_2007C_b_2007I_b_2008aq_b} \\
2007I & Ic & 2 & 180 -- 207 & \citet{2006ld_a_2007I_a, 2006ld_b_2007C_b_2007I_b_2008aq_b} \\
2007Y & Ib & 2 & 248 -- 288 & \citet{2007Y} \\
2007gr & Ic & 5 & 155 -- 388 & \citet{2007gr_a, 2007gr_b} \\
2008D & Ib & 1 & 149 & \citet{2008D_a, 2008D_b} \\
2008ax & IIb & 3 & 158 -- 359 & \citet{2008ax_a, 2008ax_b} \\
2008bo & IIb & 2 & 157 -- 217 & \citet{2007C_a_2008bo_a, 2008bo_b} \\
2009jf & Ib & 2 & 268 -- 382 & \citet{Valenti_2011_risetime_2009jf} \\
2011bm & Ic & 3 & 270 -- 303 & \citet{2011bm} \\
2011dh & IIb & 10 & 155 -- 415 & \citet{2011dh_a, 2011dh_b} \\
2011ei & IIb & 1 & 329 & \citet{2011ei} \\
2011hs & IIb & 4 & 225 -- 348 & \citet{2011hs} \\
PTF12gzk & Ic & 1 & 323 & \citet{PTF12gzk_a, PTF12gzk_b} \\
PTF12os & IIb & 2 & 139 -- 215 & \citet{PTF12os_iPTF13bvn_b}\\
iPTF13bvn & IIb & 2 & 250 -- 346 & \citet{iPTF13bvn_a, PTF12os_iPTF13bvn_b} \\
2013df & IIb & 1 & 179 & \citet{2013df} \\
2013ge & Ic & 3 & 165 -- 433 & \citet{2013ge} \\
2014eh & Ic & 1 & 236 & \citet{2014eh_a} \\
J1204 & Ib & 7 & 177 -- 271 & \citet{J1204_a, J1204_b} \\
ASASSN14az & IIb & 5 & 125 -- 190 & \citet{14az_a, 14az_b} \\
2015ah & Ib & 1 & 165 & \citet{Prentice_2019_Bigsample_2015ah} \\
2016gkg & IIb & 3 & 360 -- 483 & \citet{2016gkg} \\
2019yz & Ic & 3 & 138 -- 256 & \citet{2019yz} \\
2019odp & Ib & 3 & 147 -- 368 & \citet{2019odp} \\
2020acat & IIb & 4 & 147 -- 387 & \citet{2020acat_a, 2020acat_b} \\
2022crv & IIb & 3 & 290 -- 373 & \citet{2022crv} \\
\hline \hline

\end{tabular}

\caption{List of all SNe in our optical sample (see Section \ref{sec:observations}). The phases are given with respect to \ajc{estimated} explosion epoch. The references are not a complete list of the works on a particular SN, but will include at least the source of our used explosion epoch (first ref) and the work presenting the spectra used (second ref). If a single reference is given, it contains both of these. Note that the list here also includes all spectra from the NIR sample, presented in Table \ref{tab:NIR_sample}.} 
\label{tab:optical_sample}
\end{table*}

\begin{table*}
\begin{tabular}{lllllllllll}
\hline
SN & Type & Epoch & mag & band & Source & $A_{V\text{, tot}}$ & Method & Source & Distance & Source \\
& & [d] &  & & & & &  & [Mpc] & \\
\hline \hline
1985F & Ib & 295 & 16.9 & g & \citet{1985F_a} & 0.06 & ** & \citet{Filippenko_1985_1985Fnature} & 6.5 & CFD3 \\
\hline
1996aq & Ib & 248 &  &  & & &  & &  &  \\
\hline
1996cb & IIb & 196 & 17.22 & R & \citet{1996cb} & 0.26* &  &  & 16.8 & CFD3 \\
\hline
1997dq & Ic & 241 &  &  & &  & &  & & \\
1997dq & & 260 &  &  & &  & &  & & \\
\hline
1998bw & Ic-BL & 215 & 18.63 & V & \citet{Sollerman_2000_1998bwphot} 
& 0.57 & low-res Na I D & \citet{Patat_2001_1998bw} & 38.4 & redshift \\
1998bw & & 352 & 20.84 & V & &  & &  & & \\
\hline
2003gf & Ic & 149 &  &  & &  & &  & & \\
\hline
2004ao & Ib & 172 & 17.15 & R & \citet{2004ao_a} & 0.36 & low-res Na I D & \citet{2004ao_a} & 26.7 & NED \\
2004ao & & 208 & 17.68 & R & & & & &  &  \\
2004ao & & 247 & 18.20 & R & & & & &  &  \\
\hline
2004gt & Ic & 171 &  &  & & & & &  &  \\
\hline
2007C & Ib & 175 & 20.24 & V & \citet{Stritzinger_2018_hostreddening} & 1.44 & colour templates & \citet{Stritzinger_2018_hostreddening} & 21.0 & CFD3 \\
\hline
2007I & Ic & 180 &  &  & & & & &  &  \\
\hline
2007gr & Ic & 155 &  &  & & &  & &  &  \\
2007gr & & 170 &  &  & & & & &  &  \\
2007gr & & 387 &  &  & & & & &  &  \\
2007gr & & 388 &  &  & & & & &  &  \\
\hline
2008ax & IIb & 158 & 16.61 & R & \citet{Tsvetkov_2009_2008axphot} & 1.31 & low-res Na I D & \citet{2008ax_b} & 5.9 & NED  \\
2008ax & & 280 & 18.54 & R & &  & & &  &  \\
\hline
2009jf & Ib & 268 & & &  & &  & &  &  \\
2009jf & & 382 &  & & & &  & &  &  \\
\hline
2011bm & Ic & 299 & & &  & &  & &  &  \\
\hline
2011dh & IIb & 201 & 16.52 & R & \citet{2011dh_b} & 0.22 & high-res Na I D & \citet{2011dh_b} & 7.3 & NED \\
2011dh & & 207 & 16.61 & R & & & & &  &  \\
2011dh & & 268 & 17.77 & R & & & & &  &  \\
2011dh & & 360 & 19.53 & R & & & & &  &  \\
\hline
2011hs & IIb & 227 & 21.55 & V & \citet{2011hs} & 0.50 & low-res Na I D & \citet{2011hs} & 20.9 & CFD3 \\
\hline
2013df & IIb & 179 & 18.39 & V & \citet{2013df} & 0.33 & high-res Na I D & \citet{vanDyk_2014_2013dfextinction} & 17.9 & CFD3 \\
\hline
2014eh & Ic & 236 & 19.87 & w & \citet{2014eh_a} & 0.17* & & & 36.3 & CFD3 \\
\hline
2016gkg & IIb & 360 &  & & & &  & &  &  \\
2016gkg & & 422 &  &  & & & & &  &  \\
2016gkg & & 483 &  &  & & & & &  &  \\
\hline
2019odp & Ib & 158 & 18.60 & r & \citet{2019odp} & 0.53 & colour templates & \citet{2019odp} & 58.0 &  \\
2019odp & & 368 & 21.53 & r & & & & &  &  \\
\hline
2019yz & Ic & 256 & 20.50 & w & \citet{2019yz} & 0.31* & & & 27.0 & CFD3 \\
\hline
2022crv & IIb & 371 &  &  & &  & &  & & \\
2022crv & & 373 &  &  & &  & &  & & \\
\hline
ASASSN14az & IIb & 125 &  &  & &  && &  &  \\
ASASSN14az & & 190 &  &  & &  & &  &&  \\
\hline
J1204 & Ib & 177 & 17.2 & V & \citet{J1204_b} & 0.06* & && 15.2 & CFD3 \\
J1204 & & 182 & 17.15 & V & &  & & & & CFD3 \\
J1204 & & 203 & 17.38 & V & &  & & & & CFD3 \\
J1204 & & 243 & 18.08 & V & &  & & & & CFD3 \\
J1204 & & 261 & 18.41 & V & &  & & & & CFD3 \\
J1204 & & 266 & 18.50 & V & &  & & & & CFD3 \\
J1204 & & 271 & 18.58 & V & &  & & & & CFD3 \\
\hline
\end{tabular}

\caption{Summary information for all spectra in the NIR sample (see Section \ref{sec:observations}). Epochs are with respect to explosion date. The extinctions include extinction from the host and from the MW. For those SNe with an * in the $A_{V\text{,tot}}$ column, no estimate for the host extinction was found in the literature, and $A_{V\text{,tot}}$ thus only includes MW extinction. For SN 1985F, we assume no host extinction (see discussions in \citet{Filippenko_1985_1985Fnature} and \citet{1985F_a}). When only the epoch is given for a spectrum, this means that the spectrum was deemed not useful for a mass estimate in Section \ref{subsec:carbon_mass_estimation}, either due to missing photometry, or a poor \CIsinglet{} or \CIdoublet{} fit.  }
\label{tab:NIR_sample}
\end{table*}

\section{SUMO spectra}

\subsection{Ejecta Models}
\label{app:compmodels}

In the tables below (Tables \ref{tab:comp33} -- \ref{tab:comp80}), the detailed composition and structure of our four adopted ejecta models he3p3, he4p0, he6p0 and he8p0 are presented.


\begin{table*}
    \fontsize{7.2}{8}\selectfont
    \begin{tabular}{l|llllllllllllll}
    \hline
    & Fe/He & Si/S & O/Si/S & O/Ne/Mg & O/C & He/C$_{\text{core}}$ & He/C$_{\text{env}}$ & He/N & He/N & He/N & He/N & He/N & He/N & He/N \\ \hline \hline 
    
    $M_{\text{zone}} [M_{\odot}]$ & 0.0735 & 0.0191 & 0.0709 & 0.114 & 0.0788 & 0.0613 & 0.092 & 0.163 & 0.167 & 0.148 & 0.108 & 0.0645 & 0.0353 & 0.014 \\ 
    $V_{\text{in}}$ [km s$^{-1}$] & 0 & 2.4(3) & 2.5(3) & 3.0(3) & 3.4(3) & 3.9(3) & 4.2(3) & 4.7(3) & 5.6(3) & 6.8(3) & 8.1(3) & 9.8(3) & 1.2(4) & 1.4(4) \\ 
    $V_{\text{out}}$ [km s$^{-1}$] & 2.4(3) & 2.5(3) & 3.0(3) & 3.4(3) & 3.9(3) & 4.2(3) & 4.7(3) & 5.6(3) & 6.8(3) & 8.1(3) & 9.8(3) & 1.2(4) & 1.4(4) & 1.7(4) \\ 
    ff & 0.852 & 0.0221 & 0.0274 & 0.0441 & 0.0305 & 0.0237 & 1 & 1 & 1 & 1 & 1 & 1 & 1 & 1 \\ 
    $\rho_{\text{100d}}$ [g cm$^{-3}$] & 8.2(-16) & 8.2(-15) & 2.5(-14) & 2.5(-14) & 2.5(-14) & 2.5(-14) & 2.3(-15) & 1.5(-15) & 9.1(-16) & 4.7(-16) & 2.0(-16) & 6.9(-17) & 2.2(-17) & 5.8(-18) \\ 
    f$_{\text{clumping}}$ & 0.215 & 1.98 & 4.97 & 3.4 & 8.76 & 9.08 & 1 & 1 & 1 & 1 & 1 & 1 & 1 & 1\\
    $X_{^{56}\text{Ni}}$ & 0.662 & 0.318 & 4.6(-6) & 7.8(-7) & 7.2(-8) & 5.0(-8) & 5.0(-8) & 6.4(-9) & 6.4(-9) & 6.4(-9) & 6.4(-9) & 6.4(-9) & 6.4(-9) & 6.4(-9) \\ 
    \hline 
    $X_{\text{He}}$ & 0.202 & 1.1(-5) & 8.4(-6) & 7.0(-6) & 0.0534 & 0.832 & 0.832 & 0.985 & 0.985 & 0.985 & 0.985 & 0.985 & 0.985 & 0.985 \\ 
    $X_{\text{C}}$ & 7.0(-7) & 1.7(-6) & 2.4(-3) & 0.0124 & 0.416 & 0.135 & 0.135 & 3.6(-4) & 3.6(-4) & 3.6(-4) & 3.6(-4) & 3.6(-4) & 3.6(-4) & 3.6(-4) \\ 
    $X_{\text{N}}$ & 1.1(-6) & 5.4(-8) & 7.2(-6) & 1.9(-5) & 2.7(-5) & 2.0(-4) & 2.0(-4) & 8.6(-3) & 8.6(-3) & 8.6(-3) & 8.6(-3) & 8.6(-3) & 8.6(-3) & 8.6(-3) \\ 
    $X_{\text{O}}$ & 1.6(-5) & 1.5(-5) & 0.625 & 0.592 & 0.466 & 0.0149 & 0.0149 & 5.6(-4) & 5.6(-4) & 5.6(-4) & 5.6(-4) & 5.6(-4) & 5.6(-4) & 5.6(-4) \\ 
    $X_{\text{Ne}}$ & 1.9(-5) & 1.4(-6) & 5.5(-3) & 0.194 & 0.0458 & 0.0134 & 0.0134 & 1.2(-3) & 1.2(-3) & 1.2(-3) & 1.2(-3) & 1.2(-3) & 1.2(-3) & 1.2(-3) \\ 
    $X_{\text{Na}}$ & 5.8(-7) & 7.9(-7) & 5.7(-5) & 1.9(-3) & 1.8(-4) & 1.5(-4) & 1.5(-4) & 1.4(-4) & 1.4(-4) & 1.4(-4) & 1.4(-4) & 1.4(-4) & 1.4(-4) & 1.4(-4) \\ 
    $X_{\text{Mg}}$ & 2.4(-5) & 1.3(-4) & 0.0453 & 0.132 & 0.015 & 1.2(-3) & 1.2(-3) & 7.2(-4) & 7.2(-4) & 7.2(-4) & 7.2(-4) & 7.2(-4) & 7.2(-4) & 7.2(-4) \\ 
    $X_{\text{Si}}$ & 2.0(-4) & 0.344 & 0.247 & 0.0454 & 1.0(-3) & 8.4(-4) & 8.4(-4) & 8.2(-4) & 8.2(-4) & 8.2(-4) & 8.2(-4) & 8.2(-4) & 8.2(-4) & 8.2(-4) \\ 
    $X_{\text{S}}$ & 1.3(-4) & 0.19 & 0.0489 & 1.4(-3) & 2.5(-4) & 4.0(-4) & 4.0(-4) & 4.1(-4) & 4.1(-4) & 4.1(-4) & 4.1(-4) & 4.1(-4) & 4.1(-4) & 4.1(-4) \\ 
    $X_{\text{Ca}}$ & 1.8(-3) & 0.0328 & 2.8(-3) & 4.5(-5) & 3.2(-5) & 6.8(-5) & 6.8(-5) & 7.3(-5) & 7.3(-5) & 7.3(-5) & 7.3(-5) & 7.3(-5) & 7.3(-5) & 7.3(-5) \\ 
    $X_{\text{Fe}}$ & 1.7(-3) & 0.0607 & 4.4(-3) & 1.1(-3) & 9.5(-4) & 1.3(-3) & 1.3(-3) & 1.3(-3) & 1.3(-3) & 1.3(-3) & 1.3(-3) & 1.3(-3) & 1.3(-3) & 1.3(-3) \\ 

    \hline \hline

    \end{tabular}

    \caption{\updated{Table with the zone structure and composition for our \texttt{SUMO} input model (model $\chi = 30$) for the he3p3 model from \citet{Ertl_2020_explosions}, as described in Section \ref{subsec:ejecta_models}. \updated{In the table, f$_{\text{clumping}}$ indicates the ratio of the zone density \textit{after} macroscopic mixing divided by the density the zone would have had if left unmixed, further highlighting the "Ni-bubble effect" (see Section \ref{sec:filling_factors})}. The elemental abundances are given in mass fractions for that specific zone. Filling factor values of 1 indicate that the zone has not been macroscopically mixed (see Section \ref{sec:filling_factors}). Some zones (e.g. the He/N zone) get divided into multiple subzones of equal composition to avoid too large velocity differences within a single zone. Numbers ending on a number in brackets indicate powers of 10, e.g. 7.8(-7) = 7.8 $\times$ 10$^{-7}$.}}
    \label{tab:comp33}

\end{table*}


\begin{table*}
    \fontsize{7.2}{8}\selectfont
    \begin{tabular}{l|lllllllllllll}
    \hline
    & Fe/He & Si/S & O/Si/S & O/Ne/Mg & O/C & He/C$_{\text{core}}$ & He/C$_{\text{env}}$ & He/N & He/N & He/N & He/N & He/N & He/N \\ \hline \hline 
    
    $M_{\text{zone}} [M_{\odot}]$ & 0.0764 & 0.0334 & 0.113 & 0.316 & 0.152 & 0.0364 & 0.166 & 0.224 & 0.19 & 0.144 & 0.0808 & 0.0572 & 0.0245 \\ 
    $V_{\text{in}}$ [km s$^{-1}$] & 0 & 2.5(3) & 2.7(3) & 3.2(3) & 3.8(3) & 4.3(3) & 4.4(3) & 5.1(3) & 6.2(3) & 7.4(3) & 8.9(3) & 1.1(4) & 1.3(4) \\ 
    $V_{\text{out}}$ [km s$^{-1}$] & 2.5(3) & 2.7(3) & 3.2(3) & 3.8(3) & 4.3(3) & 4.4(3) & 5.1(3) & 6.2(3) & 7.4(3) & 8.9(3) & 1.1(4) & 1.3(4) & 1.5(4) \\ 
    ff & 0.761 & 0.0333 & 0.0377 & 0.105 & 0.0505 & 0.0121 & 1 & 1 & 1 & 1 & 1 & 1 & 1 \\ 
    $\rho_{\text{100d}}$ [g cm$^{-3}$] & 8.3(-16) & 8.3(-15) & 2.5(-14) & 2.5(-14) & 2.5(-14) & 2.5(-14) & 2.4(-15) & 1.6(-15) & 8.0(-16) & 3.5(-16) & 1.1(-16) & 4.7(-17) & 1.2(-17) \\ 
    f$_{\text{clumping}}$ & 0.233 & 1.28 & 3.91 & 2.43 & 5.36 & 8.33 & 1 & 1 & 1 & 1 & 1 & 1 & 1 \\
    $X_{^{56}\text{Ni}}$ & 0.681 & 0.277 & 2.1(-6) & 7.0(-7) & 7.8(-8) & 3.4(-8) & 3.4(-8) & 1.9(-8) & 1.9(-8) & 1.9(-8) & 1.9(-8) & 1.9(-8) & 1.9(-8) \\ 
    \hline 
    $X_{\text{He}}$ & 0.184 & 9.5(-6) & 6.1(-6) & 5.0(-6) & 0.0483 & 0.813 & 0.813 & 0.985 & 0.985 & 0.985 & 0.985 & 0.985 & 0.985 \\ 
    $X_{\text{C}}$ & 7.4(-7) & 4.8(-6) & 2.5(-3) & 0.0192 & 0.396 & 0.148 & 0.148 & 3.5(-4) & 3.5(-4) & 3.5(-4) & 3.5(-4) & 3.5(-4) & 3.5(-4) \\ 
    $X_{\text{N}}$ & 7.1(-7) & 1.1(-7) & 5.4(-6) & 3.1(-5) & 2.3(-5) & 1.2(-4) & 1.2(-4) & 8.6(-3) & 8.6(-3) & 8.6(-3) & 8.6(-3) & 8.6(-3) & 8.6(-3) \\ 
    $X_{\text{O}}$ & 1.5(-5) & 1.3(-4) & 0.583 & 0.514 & 0.511 & 0.0205 & 0.0205 & 6.2(-4) & 6.2(-4) & 6.2(-4) & 6.2(-4) & 6.2(-4) & 6.2(-4) \\ 
    $X_{\text{Ne}}$ & 1.8(-5) & 2.4(-6) & 5.6(-3) & 0.337 & 0.0337 & 0.0142 & 0.0142 & 1.2(-3) & 1.2(-3) & 1.2(-3) & 1.2(-3) & 1.2(-3) & 1.2(-3) \\ 
    $X_{\text{Na}}$ & 4.3(-7) & 7.9(-7) & 5.7(-5) & 7.9(-3) & 1.8(-4) & 1.4(-4) & 1.4(-4) & 1.4(-4) & 1.4(-4) & 1.4(-4) & 1.4(-4) & 1.4(-4) & 1.4(-4) \\ 
    $X_{\text{Mg}}$ & 2.2(-5) & 1.4(-4) & 0.0354 & 0.1 & 7.9(-3) & 9.7(-4) & 9.7(-4) & 7.2(-4) & 7.2(-4) & 7.2(-4) & 7.2(-4) & 7.2(-4) & 7.2(-4) \\ 
    $X_{\text{Si}}$ & 2.0(-4) & 0.386 & 0.286 & 0.0107 & 9.4(-4) & 8.3(-4) & 8.3(-4) & 8.2(-4) & 8.2(-4) & 8.2(-4) & 8.2(-4) & 8.2(-4) & 8.2(-4) \\ 
    $X_{\text{S}}$ & 1.2(-4) & 0.187 & 0.0614 & 4.8(-4) & 2.9(-4) & 4.0(-4) & 4.0(-4) & 4.1(-4) & 4.1(-4) & 4.1(-4) & 4.1(-4) & 4.1(-4) & 4.1(-4) \\ 
    $X_{\text{Ca}}$ & 1.8(-3) & 0.0269 & 1.9(-3) & 4.7(-5) & 4.1(-5) & 7.0(-5) & 7.0(-5) & 7.3(-5) & 7.3(-5) & 7.3(-5) & 7.3(-5) & 7.3(-5) & 7.3(-5) \\ 
    $X_{\text{Fe}}$ & 1.5(-3) & 0.0705 & 3.3(-3) & 1.1(-3) & 1.2(-3) & 1.3(-3) & 1.3(-3) & 1.3(-3) & 1.3(-3) & 1.3(-3) & 1.3(-3) & 1.3(-3) & 1.3(-3) \\ 

    \hline \hline

    \end{tabular}

    \caption{\updated{Same as table \ref{tab:comp33}, but now for the he4p0 model.}}
    \label{tab:comp40}

\end{table*}


\begin{table*}
    \fontsize{7.2}{8}\selectfont
    \begin{tabular}{l|llllllllllll}
    \hline
    & Fe/He & Si/S & O/Si/S & O/Ne/Mg & O/Ne/Mg & O/C & He/C$_{\text{core}}$ & He/C$_{\text{env}}$ & He/N & He/N & He/N & He/N \\ \hline \hline 
    
    $M_{\text{zone}} [M_{\odot}]$ & 0.115 & 0.0328 & 0.167 & 0.348 & 0.75 & 0.308 & 0.0448 & 0.403 & 0.272 & 0.172 & 0.116 & 0.0606 \\ 
    $V_{\text{in}}$ [km s$^{-1}$] & 0 & 2.2(3) & 2.4(3) & 3.1(3) & 3.7(3) & 4.8(3) & 5.5(3) & 5.6(3) & 6.9(3) & 8.3(3) & 1.0(4) & 1.2(4) \\ 
    $V_{\text{out}}$ [km s$^{-1}$] & 2.2(3) & 2.4(3) & 3.1(3) & 3.7(3) & 4.8(3) & 5.5(3) & 5.6(3) & 6.9(3) & 8.3(3) & 1.0(4) & 1.2(4) & 1.5(4) \\ 
    ff & 0.668 & 0.019 & 0.0323 & 0.0674 & 0.145 & 0.0596 & 8.6(-3) & 1 & 1 & 1 & 1 & 1 \\ 
    $\rho_{\text{100d}}$ [g cm$^{-3}$] & 7.0(-16) & 7.0(-15) & 2.1(-14) & 2.1(-14) & 2.1(-14) & 2.1(-14) & 2.1(-14) & 1.9(-15) & 8.1(-16) & 3.0(-16) & 1.2(-16) & 2.8(-17) \\ 
    f$_{\text{clumping}}$ & 9.68(-2) & 0.899 & 2.72 & 1.83 & 2.24 & 5.30 & 6.91 & 1 & 1 & 1 & 1 & 1 \\
    $X_{^{56}\text{Ni}}$ & 0.67 & 0.207 & 4.4(-6) & 2.9(-7) & 2.9(-7) & 8.0(-8) & 7.8(-8) & 7.8(-8) & 3.2(-8) & 3.2(-8) & 3.2(-8) & 3.2(-8) \\ 
    \hline 
    $X_{\text{He}}$ & 0.211 & 1.0(-5) & 4.7(-6) & 2.9(-6) & 2.9(-6) & 0.0502 & 0.584 & 0.584 & 0.985 & 0.985 & 0.985 & 0.985 \\ 
    $X_{\text{C}}$ & 1.1(-6) & 2.2(-6) & 2.0(-3) & 9.8(-3) & 9.8(-3) & 0.387 & 0.274 & 0.274 & 2.8(-4) & 2.8(-4) & 2.8(-4) & 2.8(-4) \\ 
    $X_{\text{N}}$ & 7.9(-7) & 4.5(-8) & 5.5(-6) & 1.4(-5) & 1.4(-5) & 1.8(-5) & 6.2(-4) & 6.2(-4) & 8.9(-3) & 8.9(-3) & 8.9(-3) & 8.9(-3) \\ 
    $X_{\text{O}}$ & 1.7(-5) & 1.9(-5) & 0.668 & 0.582 & 0.582 & 0.539 & 0.124 & 0.124 & 2.7(-4) & 2.7(-4) & 2.7(-4) & 2.7(-4) \\ 
    $X_{\text{Ne}}$ & 1.7(-5) & 1.3(-6) & 6.2(-3) & 0.277 & 0.277 & 0.0164 & 0.0141 & 0.0141 & 1.1(-3) & 1.1(-3) & 1.1(-3) & 1.1(-3) \\ 
    $X_{\text{Na}}$ & 5.1(-7) & 6.6(-7) & 4.6(-5) & 3.8(-3) & 3.8(-3) & 1.7(-4) & 1.4(-4) & 1.4(-4) & 1.4(-4) & 1.4(-4) & 1.4(-4) & 1.4(-4) \\ 
    $X_{\text{Mg}}$ & 2.0(-5) & 1.0(-4) & 0.0397 & 0.103 & 0.103 & 4.1(-3) & 7.8(-4) & 7.8(-4) & 7.2(-4) & 7.2(-4) & 7.2(-4) & 7.2(-4) \\ 
    $X_{\text{Si}}$ & 2.7(-4) & 0.405 & 0.217 & 0.0127 & 0.0127 & 8.8(-4) & 8.2(-4) & 8.2(-4) & 8.2(-4) & 8.2(-4) & 8.2(-4) & 8.2(-4) \\ 
    $X_{\text{S}}$ & 2.1(-4) & 0.235 & 0.0439 & 4.0(-4) & 4.0(-4) & 3.3(-4) & 4.1(-4) & 4.1(-4) & 4.1(-4) & 4.1(-4) & 4.1(-4) & 4.1(-4) \\ 
    $X_{\text{Ca}}$ & 2.1(-3) & 0.0398 & 2.7(-3) & 4.1(-5) & 4.1(-5) & 5.3(-5) & 7.2(-5) & 7.2(-5) & 7.4(-5) & 7.4(-5) & 7.4(-5) & 7.4(-5) \\ 
    $X_{\text{Fe}}$ & 1.7(-3) & 0.054 & 3.2(-3) & 1.0(-3) & 1.0(-3) & 1.1(-3) & 1.3(-3) & 1.3(-3) & 1.3(-3) & 1.3(-3) & 1.3(-3) & 1.3(-3) \\ 

    \hline \hline

    \end{tabular}

    \caption{\updated{Same as table \ref{tab:comp33}, but now for the he6p0 model.}}
    \label{tab:comp60}

\end{table*}


\begin{table*}
    \fontsize{7.2}{8}\selectfont
    \begin{tabular}{l|llllllllllllll}
    \hline
    & Fe/He & Si/S & O/Si/S & O/Ne/Mg & O/Ne/Mg & O/Ne/Mg & O/Ne/Mg & O/C & He/C$_{\text{core}}$ & He/C$_{\text{env}}$ & He/C$_{\text{env}}$ & He/N & He/N & He/N \\ \hline \hline 
    
    $M_{\text{zone}} [M_{\odot}]$ & 0.083 & 0.0336 & 0.142 & 0.381 & 0.555 & 0.645 & 0.398 & 0.36 & 0.087 & 0.381 & 0.397 & 0.225 & 0.147 & 0.0833 \\ 
    $V_{\text{in}}$ [km s$^{-1}$] & 0 & 1.3(3) & 1.4(3) & 1.7(3) & 2.0(3) & 2.4(3) & 2.9(3) & 3.3(3) & 3.8(3) & 4.0(3) & 4.8(3) & 5.9(3) & 7.0(3) & 8.5(3) \\ 
    $V_{\text{out}}$ [km s$^{-1}$] & 1.3(3) & 1.4(3) & 1.7(3) & 2.0(3) & 2.4(3) & 2.9(3) & 3.3(3) & 3.8(3) & 4.0(3) & 4.8(3) & 5.9(3) & 7.0(3) & 8.5(3) & 1.1(4) \\ 
    ff & 0.483 & 0.0195 & 0.0276 & 0.0738 & 0.107 & 0.125 & 0.0771 & 0.0698 & 0.0169 & 1 & 1 & 1 & 1 & 1 \\ 
    $\rho_{\text{100d}}$ [g cm$^{-3}$] & 2.0(-15) & 2.0(-14) & 5.9(-14) & 5.9(-14) & 5.9(-14) & 5.9(-14) & 5.9(-14) & 5.9(-14) & 5.9(-14) & 6.0(-15) & 3.1(-15) & 1.1(-15) & 4.2(-16) & 9.4(-17) \\ 
    f$_{\text{clumping}}$ & 7.61(-2) & 0.330 & 1.35 & 0.794 & 0.941 & 1.4 & 2.11 & 4.45 & 6.34 1 & 1 & 1 & 1 & 1 \\
    $X_{^{56}\text{Ni}}$ & 0.649 & 0.213 & 1.5(-6) & 2.0(-7) & 2.0(-7) & 2.0(-7) & 2.0(-7) & 6.8(-8) & 6.9(-8) & 6.9(-8) & 6.9(-8) & 2.2(-8) & 2.2(-8) & 2.2(-8) \\ 
    \hline 
    $X_{\text{He}}$ & 0.229 & 1.3(-5) & 4.5(-6) & 3.4(-6) & 3.4(-6) & 3.4(-6) & 3.4(-6) & 0.017 & 0.377 & 0.377 & 0.377 & 0.985 & 0.985 & 0.985 \\ 
    $X_{\text{C}}$ & 1.5(-6) & 9.9(-6) & 2.9(-3) & 0.0179 & 0.0179 & 0.0179 & 0.0179 & 0.31 & 0.391 & 0.391 & 0.391 & 3.6(-4) & 3.6(-4) & 3.6(-4) \\ 
    $X_{\text{N}}$ & 5.5(-7) & 1.4(-7) & 6.5(-6) & 2.4(-5) & 2.4(-5) & 2.4(-5) & 2.4(-5) & 1.5(-5) & 8.5(-4) & 8.5(-4) & 8.5(-4) & 8.9(-3) & 8.9(-3) & 8.9(-3) \\ 
    $X_{\text{O}}$ & 2.0(-5) & 6.4(-4) & 0.7 & 0.604 & 0.604 & 0.604 & 0.604 & 0.648 & 0.214 & 0.214 & 0.214 & 2.7(-4) & 2.7(-4) & 2.7(-4) \\ 
    $X_{\text{Ne}}$ & 1.9(-5) & 5.8(-6) & 6.9(-3) & 0.287 & 0.287 & 0.287 & 0.287 & 0.0162 & 0.0137 & 0.0137 & 0.0137 & 1.1(-3) & 1.1(-3) & 1.1(-3) \\ 
    $X_{\text{Na}}$ & 6.0(-7) & 6.9(-7) & 6.0(-5) & 7.8(-3) & 7.8(-3) & 7.8(-3) & 7.8(-3) & 1.8(-4) & 1.4(-4) & 1.4(-4) & 1.4(-4) & 1.4(-4) & 1.4(-4) & 1.4(-4) \\ 
    $X_{\text{Mg}}$ & 1.8(-5) & 1.3(-4) & 0.0365 & 0.0699 & 0.0699 & 0.0699 & 0.0699 & 5.8(-3) & 9.8(-4) & 9.8(-4) & 9.8(-4) & 7.2(-4) & 7.2(-4) & 7.2(-4) \\ 
    $X_{\text{Si}}$ & 3.2(-4) & 0.402 & 0.196 & 5.4(-3) & 5.4(-3) & 5.4(-3) & 5.4(-3) & 9.0(-4) & 8.3(-4) & 8.3(-4) & 8.3(-4) & 8.2(-4) & 8.2(-4) & 8.2(-4) \\ 
    $X_{\text{S}}$ & 2.6(-4) & 0.233 & 0.0377 & 3.1(-4) & 3.1(-4) & 3.1(-4) & 3.1(-4) & 2.9(-4) & 4.0(-4) & 4.0(-4) & 4.0(-4) & 4.1(-4) & 4.1(-4) & 4.1(-4) \\ 
    $X_{\text{Ca}}$ & 2.6(-3) & 0.0396 & 1.9(-3) & 3.8(-5) & 3.8(-5) & 3.8(-5) & 3.8(-5) & 4.5(-5) & 7.1(-5) & 7.1(-5) & 7.1(-5) & 7.4(-5) & 7.4(-5) & 7.4(-5) \\ 
    $X_{\text{Fe}}$ & 1.8(-3) & 0.0538 & 2.2(-3) & 8.9(-4) & 8.9(-4) & 8.9(-4) & 8.9(-4) & 1.0(-3) & 1.3(-3) & 1.3(-3) & 1.3(-3) & 1.3(-3) & 1.3(-3) & 1.3(-3) \\

    \hline \hline

    \end{tabular}

    \caption{\updated{Same as table \ref{tab:comp33}, but now for the he8p0 model.}}
    \label{tab:comp80}

\end{table*}

\subsection{Synthetic Spectra}

\begin{figure*}
    \centering
    \includegraphics[width=0.99\textwidth]{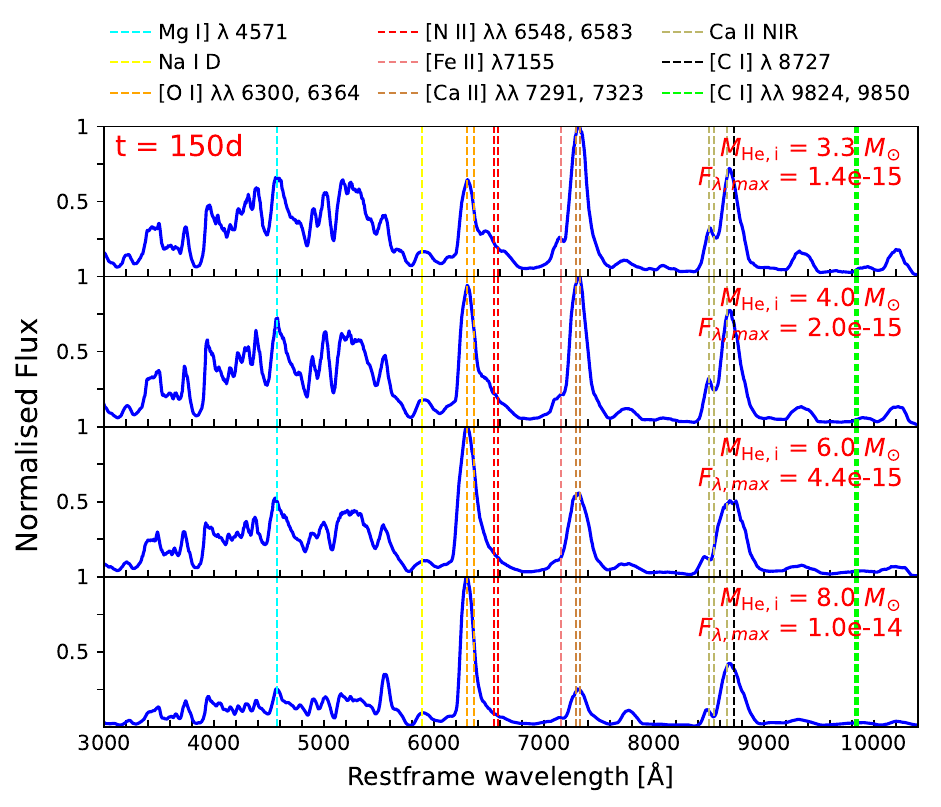}
    \caption{Synthetic \SUMOcodename{} spectra for our $\chi = 30$ ejecta models at 150d post explosion. Each spectrum is normalised to the peak flux in the shown wavelength range. \updated{The maximum flux $F_{\lambda, max}$ in each spectrum (i.e. where normalised flux = 1) is given in units of \fluxunit. }  Vertical dashed lines indicate the locations of the most prominent emission lines.}
    \label{fig:spectra_150d}
\end{figure*}

\begin{figure*}
    \centering
    \includegraphics[width=0.99\textwidth]{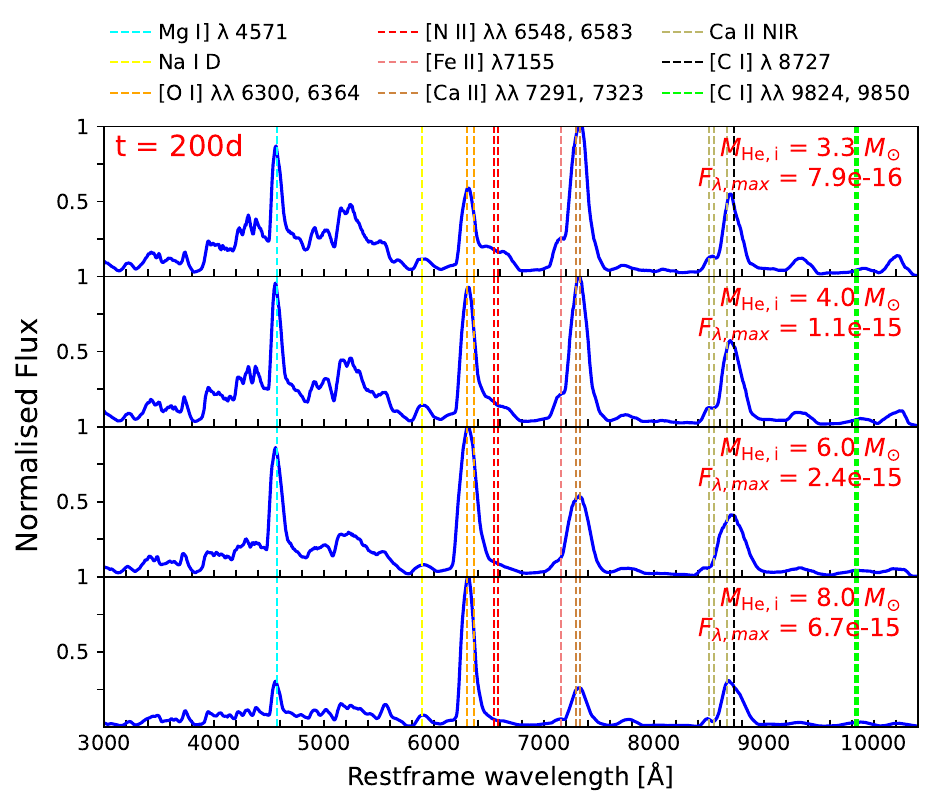}
    \caption{Same as Figure \ref{fig:spectra_150d}, but now at 200d post explosion.}
    \label{fig:spectra_200d}
\end{figure*}

\begin{figure*}
    \centering
    \includegraphics[width=0.99\textwidth]{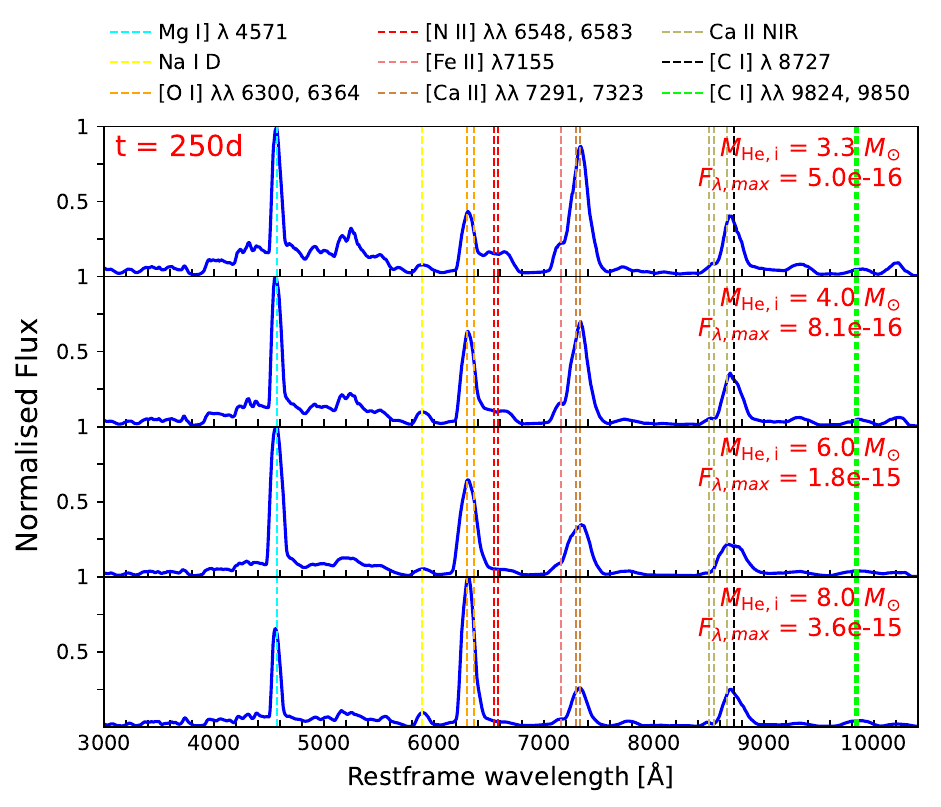}
    \caption{Same as Figure \ref{fig:spectra_150d}, but now at 250d post explosion.}
    \label{fig:spectra_250d}
\end{figure*}

\begin{figure*}
    \centering
    \includegraphics[width=0.99\textwidth]{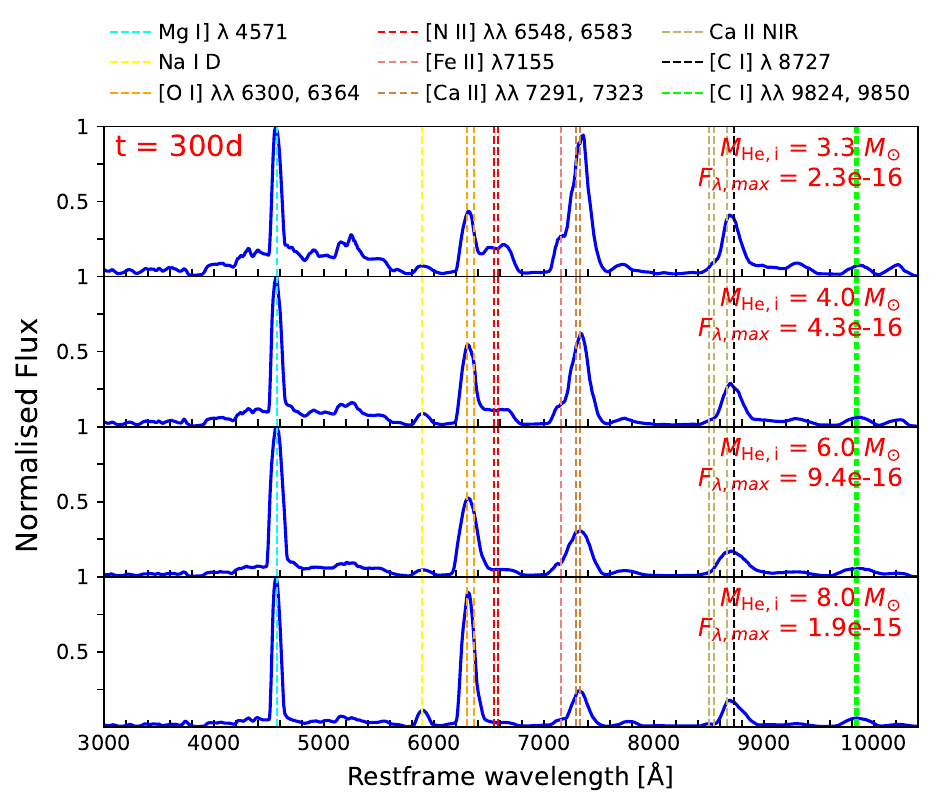}
    \caption{Same as Figure \ref{fig:spectra_150d}, but now at 300d post explosion.}
    \label{fig:spectra_300d}
\end{figure*}

\begin{figure*}
    \centering
    \includegraphics[width=0.99\textwidth]{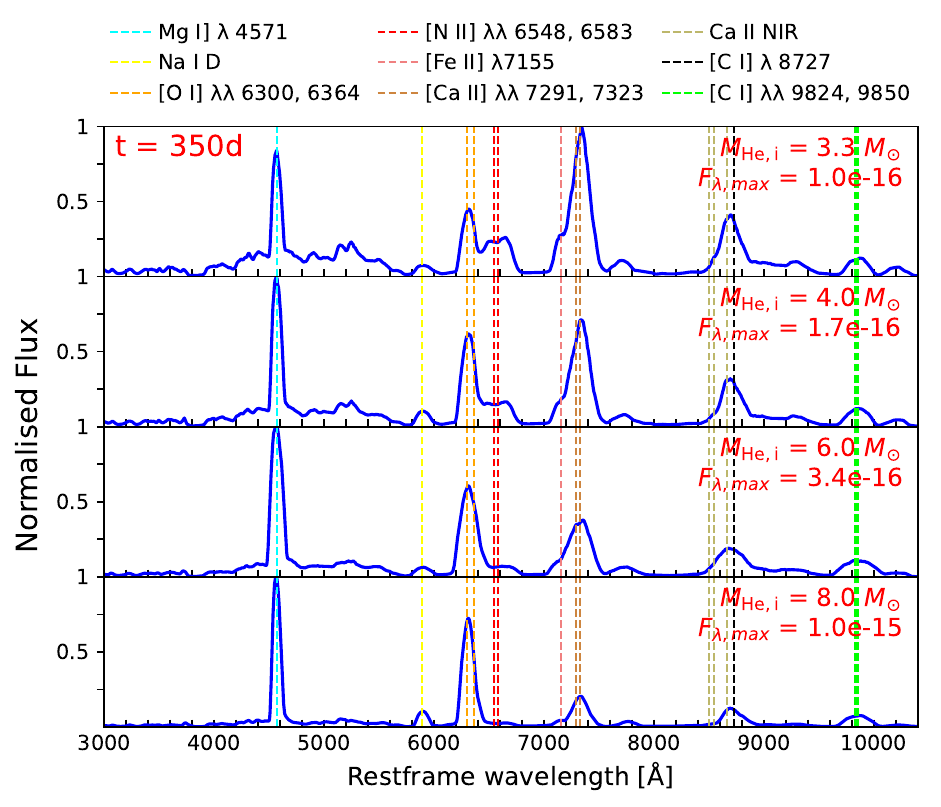}
    \caption{Same as Figure \ref{fig:spectra_150d}, but now at 350d post explosion.}
    \label{fig:spectra_350d}
\end{figure*}

\begin{figure*}
    \centering
    \includegraphics[width=0.99\textwidth]{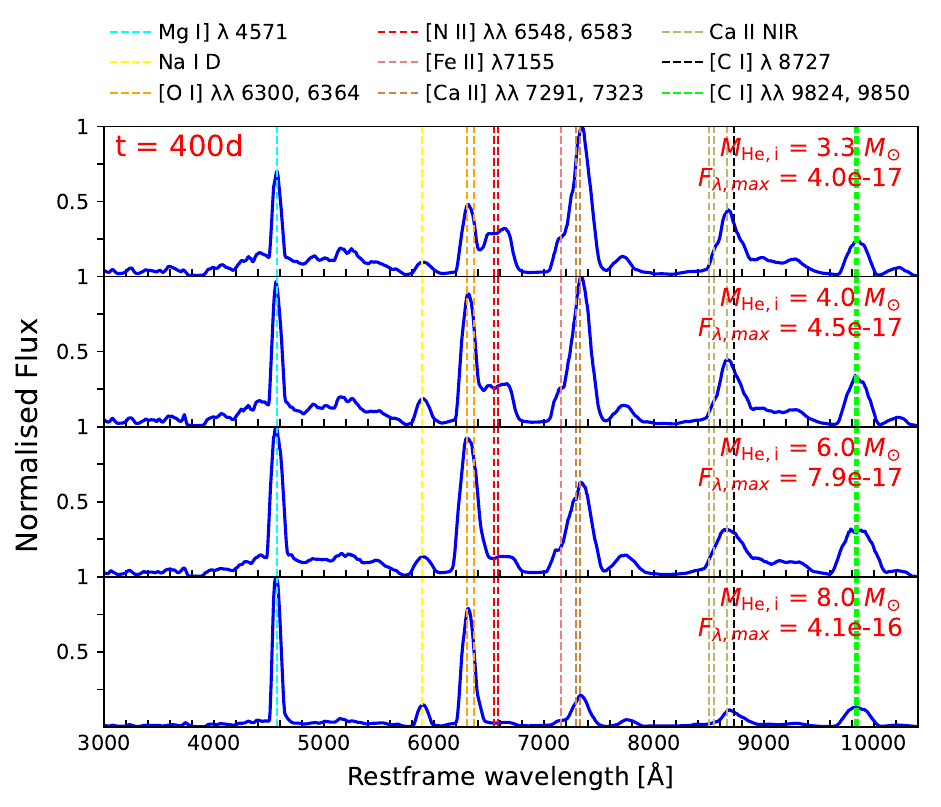}
    \caption{Same as Figure \ref{fig:spectra_150d}, but now at 400d post explosion.}
    \label{fig:spectra_400d}
\end{figure*}

Here we present the \SUMOcodename{} spectra for the $\chi = 30$ set, separated by epoch in Figures \ref{fig:spectra_150d} --  \ref{fig:spectra_400d}.

\end{document}